\documentclass[useAMS,usenatbib]{mn2e}
\usepackage[letterpaper,total={17.8cm,24.0cm},centering]{geometry}

\usepackage{graphicx} \usepackage{aas_macros} \usepackage{lscape}
\usepackage{amsmath}

\newcommand{\gtsim}{\mbox{{\raisebox{-0.4ex}{$\stackrel{>}{{\scriptstyle\sim}}$}}}}

\graphicspath{{plots_april/}}

\title[Clustered star formation in major mergers.]{Beyond the nuclear starburst? Clustered star formation in major mergers.} 
\author[L. C. Powell et al]{Leila C. Powell$^{1,2}$\thanks{E-mail: lpowell@mpe.mpg.de}, Frederic Bournaud$^{1}$, Damien Chapon$^{1}$ and Romain Teyssier$^{1,3}$\\ 
$^{1}$CEA Saclay, DSM/IRFU/SAP, Orme des Merisiers, 91191 Gif-sur-Yvette Cedex, France\\
$^{2}$Max Planck Institute for extraterrestrial Physics, PO Box 1312, Giessenbachstr., 85741 Garching, Germany\\
$^{3}$Institute for Theoretical Physics, University of Z{\" u}rich, Winterthurestrasse 190, CH-8057 Z{\" u}rich, Switzerland}

\begin{document}

\date{Accepted . Received ; in original form }

\pagerange{\pageref{firstpage}--\pageref{lastpage}} \pubyear{2011}

\maketitle

\label{firstpage}

\begin{abstract}
 
Recent simulation work has successfully captured the formation of the star clusters that have been observed in merging galaxies. These studies, however, tend to focus on studying extreme starbursts, such as the Antennae galaxies.  We aim to establish whether there is something special occurring in these extreme systems or whether the mechanism for cluster formation is present in all mergers to a greater or lesser degree. We undertake a general study of merger-induced star formation in a sample of $5$ pc resolution adaptive mesh refinement simulations of low redshift equal-mass mergers with randomly-chosen orbital parameters. We find that there is an enhanced mass fraction of very dense gas that appears as the gas density probability density function evolves during the merger. This finding has implications for the interpretation of some observations; a larger mass fraction of dense gas could account for the enhanced HCN/CO ratios seen in ULIRGs and predicts that $\alpha_{\rm CO}$ is lower in mergers, as for a given mass of ${\rm H}_2$, CO emission will increase in a denser environment. We also find that as the star formation rate increases, there is a correlated peak in the velocity dispersion of the gas, which we attribute to increasing turbulence driven by the interaction itself. Star formation tends to be clumpy: in some cases there is {\it extended} clumpy star formation, but even when star formation is concentrated within the inner kpc (i.e. what may be considered a nuclear starburst) it still often has a clumpy, rather than a smooth, distribution. We find no strong evidence for a clear bimodality in the Kennicutt-Schmidt relation for the average mergers simulated here. Instead, they are typically somewhat offset above the predicted quiescent relation during their starbursts.

\end{abstract}


\section{Introduction}

Galaxy-galaxy mergers are a key ingredient in the current hierarchical framework for structure formation. These interactions are theorised to be responsible for many of the observed stages of galaxy evolution: the transformation of spirals into ellipticals \citep[][]{Toomre1977,Schweizer1982}, the growth of bulges \citep[e.g.][]{BournaudJogCombes2005}, the destruction of discs  \citep{HopkinsCoxYounger2009,ScannapiecoWhiteSpringel2009,StewartBullockWechsler2009}, the creation of dwarf galaxies \citep{DucMirabel1994,ElmegreenKaufmanThomasson1993,MirabelDottoriLutz1992} and the extreme star formation rates (SFRs) in some local galaxy populations \citep[e.g.][]{BarnesHernquist1991, MihosHernquist1994} \citep[see][for a review]{BarnesHernquist1992}. 

While it is clear from observations that galaxy-galaxy mergers occur frequently in nature \citep{ConseliceBershadyDickinson2003}, their exact role and the {\it extent} of their influence on galaxy evolution is still debated. The contribution of starbursts to the global budget of stars formed at $z\le 2$ has been placed at $80\%$ by \citet{ElbazCesarsky2003}, but at only $10\%$ for merger-induced star formation at $z\le 1$ by \cite{RobainaBellSkelton2009}. Similarly, \citet{RodighieroDaddiBaronchelli2011} find only $10\%$ of the cosmic SFR density at $z\sim2$ comes from starbursting galaxies. Simulations have also demonstrated that most baryonic mass is accreted not via major mergers, but rather via cold flows or minor mergers \citep[e.g.][]{KeresKatzWeinberg2005, DekelBirnboimEngel2009, BrooksGovernatoQuinn2009}. Furthermore, due in some part to ongoing improvements in  hydrodynamical simulations, alternative explanations for many supposed  `merger-induced' features have recently been proposed. For example, bulge formation via clump migration \citep[][]{ceverino_dekel_bournaud2010} and disc reformation resulting from cold accretion \citep[e.g.][]{highzmergers}. 

Due to the complex dynamics at play during mergers, simulations have proved to be an invaluable tool in understanding the underlying physics. As first demonstrated with merger simulations by \citet{BarnesHernquist1991}, tidal torques on the galaxies, due to their interaction, drive material inwards in the central regions, resulting in a high concentration of gas at the nucleus. This translates into a significant increase in the SFR and a classic `nuclear starburst'. This description fits well with observations of LIRGs and ULIRGs, centrally-concentrated starbursting galaxies \citep{SandersSoiferElias1988,DucMirabelMaza1997}. We note, however, that the highest IR luminosities in ULIRGs are often attributed to AGN activity, either instead of, or as well as, a merger-induced starburst \citep{YuanKewleySanders2010}.

There is also mounting evidence, however, for a clustered component of merger-induced star formation. LBGs are irregular and are therefore often proposed to be merging systems. \citet{OverzierHeckmanKauffmann2008} find star formation in local analogs of LBGs is dominated by unresolved `super starburst regions', which they propose consist of star clusters. This result leads them to suggest that star formation in high-redshift LBGs may also be clustered, but not resolved in observations and, indeed, star-forming `knots' are revealed in gravitationally lensed high-redshift galaxies \citep[e.g.][]{FranxIllingworthKelson1997}. 

An important difference between a nuclear starburst and clustered star formation is that the latter is not confined to the galactic centre and the induced star formation can potentially occur over an extended region. A well-studied example of this extended, clustered star formation is the Antennae system \citep{WhitmoreSchweizer1995}. In this case, the majority of the star formation is outside the nuclei \citep{WangFazioAshby2004}. There are also merging systems in which significant star formation occurs in tidal features, such as clumps in tidal tails \citep[e.g.][]{2008AJ....135.2406S} and tidal dwarf galaxies \citep[e.g.][]{WeilbacherDucFritze-v.Alvensleben2000, HancockSmithStruck2009}.

There are further indications that the physical processes during a merger that produce the starburst are not limited to an increase of the gas surface density, $\Sigma_{\rm gas}$, due to global gas compression (this process  and the way it gives rise to the nuclear starburst are well understood). If this were the only mechanism at work, we would expect that the ratio $\Sigma_{\rm SFR}/\Sigma_{\rm gas}$, where $\Sigma_{\rm SFR}$ is the SFR surface density, would not diverge from the value observed for quiescent disc galaxies. Recent work by  \citet{daddi_etal_2010} and \citet{GenzelTacconiGracia-Carpio2010} indicates, however, that there could be a bimodality in the Kennicutt-Schmidt relation ($\Sigma_{\rm SFR} - \Sigma_{\rm gas}$), with starbursting discs positioned about a dex above quiescent discs. This suggests that merger-induced starbursts may involve more complex physical processes.

An additional mechanism is required to explain both this deviation in the Kennicutt-Schmidt law for starbursts and the observed merger-induced {\it clustered} star formation. Much simulation work has been done with the aim of reproducing (and understanding) the extended stellar distribution observed in some merging systems, including testing multiple star formation recipes \citep[e.g.][shock-induced star formation]{barnes2004}. More recently, \citet{antennae_ramses} have investigated clustered star formation in the Antennae galaxies and suggested this is also related to bimodality in the Kennicutt-Schmidt relation. 

Often, one of the most important considerations in hydrodynamical simulations (and one of the most severe limitations) is the resolution. Historically, most idealised merger simulations were performed with a stabilised interstellar medium (ISM) (limited resolution equates to a limited minimum gas temperature) \citep[e.g.][]{Di-MatteoBournaudMartig2008}. While this is perfectly adequate for studying the global response of the gas to tidal torques, this approach will never capture the multiphase nature of the ISM and star formation will inevitably be smoothly distributed over the disc. Any clustered star formation that may have occurred will simply be missed. This is not just important for merger-induced star formation; if star formation is clumpy then the pre-merger discs will have a different structure and this can affect the course of the merger.

This issue was examined in detail for high redshift mergers (which are extremely clumpy) by \citet{highzmergers} who show that having clumpy, rather than smooth, pre-merger discs affects everything from the SFR to the remnant properties. \citet{antennae_ramses} demonstrate that  increasing the spatial resolution in simulations of the Antennae system increases the SFR significantly as the clustered star formation that is then resolved adds to the existing nuclear starburst (however they do not resolve clumps in the pre-merger discs and do not include feedback). While high redshift galaxies are more obviously clumpy (in the sense that they have fewer, more massive clumps), low redshift discs are still multiphase, with cool clouds embedded in a warmer medium.  Since it is in the clumps/clouds that star formation occurs it is vital to resolve the overdensities of at least the most massive clumps/clouds properly so that the computed SFR is correct. At high redshift, it is `easier' to resolve the most massive clumps, requiring only $\sim 100$pc, whereas the less massive low redshift clouds require resolutions of a few pc. Therefore, resolution is still a very important issue when studying low redshift mergers, which is the focus of this work.

In this paper, we use a set of high resolution ($\approx 5$ pc) idealised adaptive mesh refinement (AMR) simulations to investigate merger-induced star formation in major mergers. The orbital parameters in our sample are chosen to be `average', such that we can investigate how star formation proceeds in general in the galaxy population, rather than focusing on specific observed systems that exhibit particularly striking stellar morphologies. Our main goals are:

\begin{itemize}
\item To measure the changes in the gas properties (fraction of dense gas, the velocity dispersion etc.) during average mergers, in order to `observe', at high resolution, the processes which enhance star formation. 
\item To look for signatures, in average mergers, of the process that can produce Super Star Clusters in some more extreme examples of interacting systems (see preceding discussion).
\item  To explore how this process differs from the classic nuclear starburst picture and how the combination of these two mechanisms affects the star formation.
\item To revisit the interpretation of some recent observations of starbursts based on our findings, focusing on the question of bimodality in the Kennicutt-Schmidt relation.
\end{itemize}

Briefly, our main findings are as follows. The interactions result in a significant increase in the mass fraction of very dense gas leading to enhanced star formation. We find that the majority of mergers in our sample have a non-negligible component of extended ($>1$ kpc), clustered star formation and in some cases this accounts for the majority of the star formation at the early stages of the starburst. In all cases, as the merger progresses, the star formation becomes increasingly centrally concentrated, resembling, in terms of size, a classic nuclear starburst. The gas distribution within the central region is often still clumpy, however, somewhat in contrast to the classic picture. We do not find that the starbursting galaxies in our sample lie on a separate Kennicutt-Schmidt sequence as proposed in \citet{daddi_etal_2010} and \citet{GenzelTacconiGracia-Carpio2010}, but rather that they typically lie somewhere between this and the K-S sequence for quiescent discs. We note, however, that \citet{daddi_etal_2010} and \citet{GenzelTacconiGracia-Carpio2010} select only the most extreme starbursts for their analysis.

\section{The sample of merging galaxies}\label{sec:sim}

We simulate a sample of $5$ equal mass mergers in live dark matter haloes (using the same initial morphology for both galaxies, approximately an Sb spiral) and have also evolved one of the galaxies in isolation. In particular, we use the isolated galaxy to calibrate our choice of star formation and feedback parameters, such that the discs have a reasonable SFR. The isolated disc also provides a useful point of comparison when we are measuring properties of our merging galaxies, allowing us to isolate the impact of the merger process. The exact values of properties are always somewhat dependent on choices of sub-grid recipes, but we can measure the relative change between the isolated and merging galaxies.

\begin{figure*}
\includegraphics[width=\textwidth]{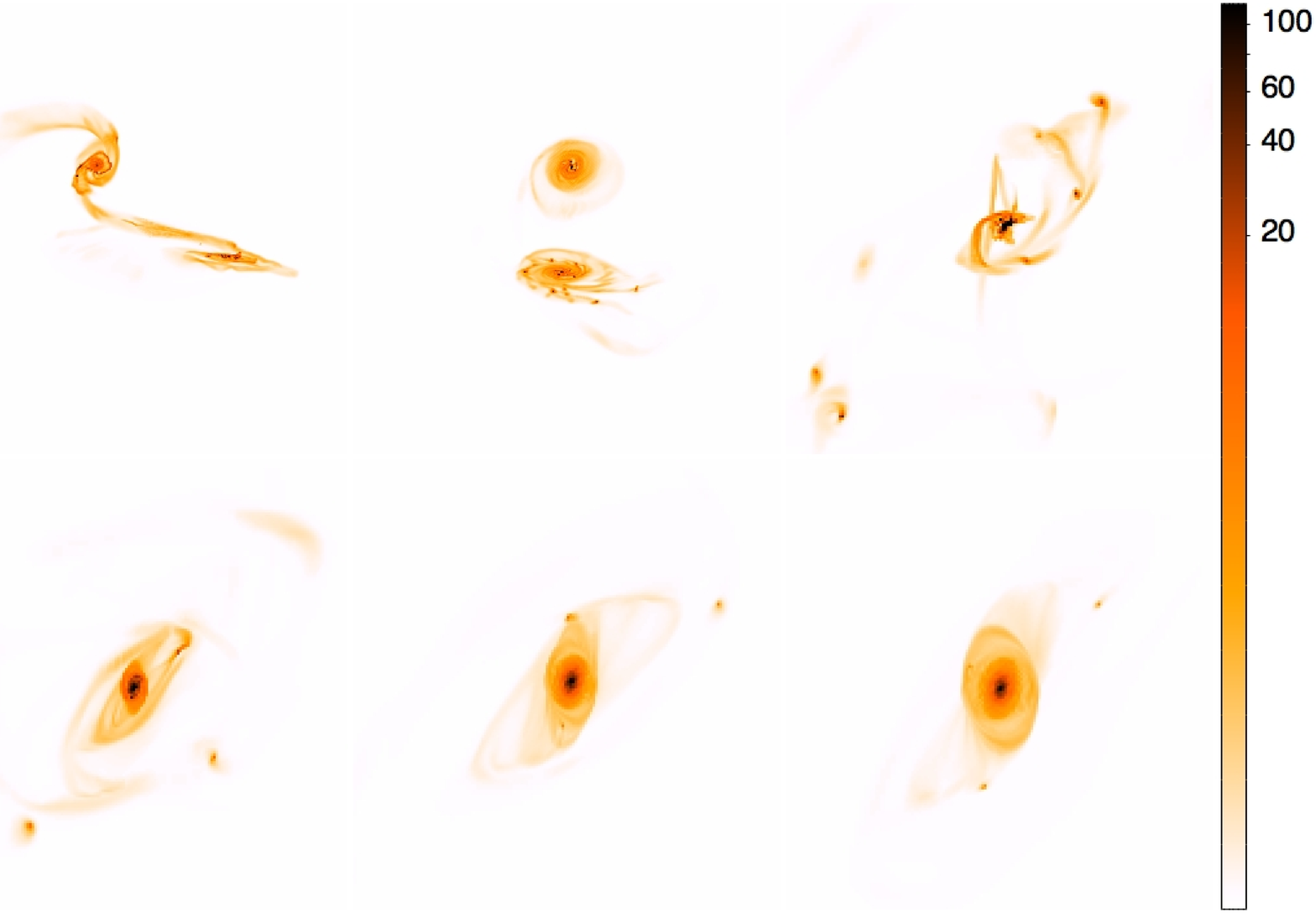}
\vskip 0.5cm
\includegraphics[width=\textwidth]{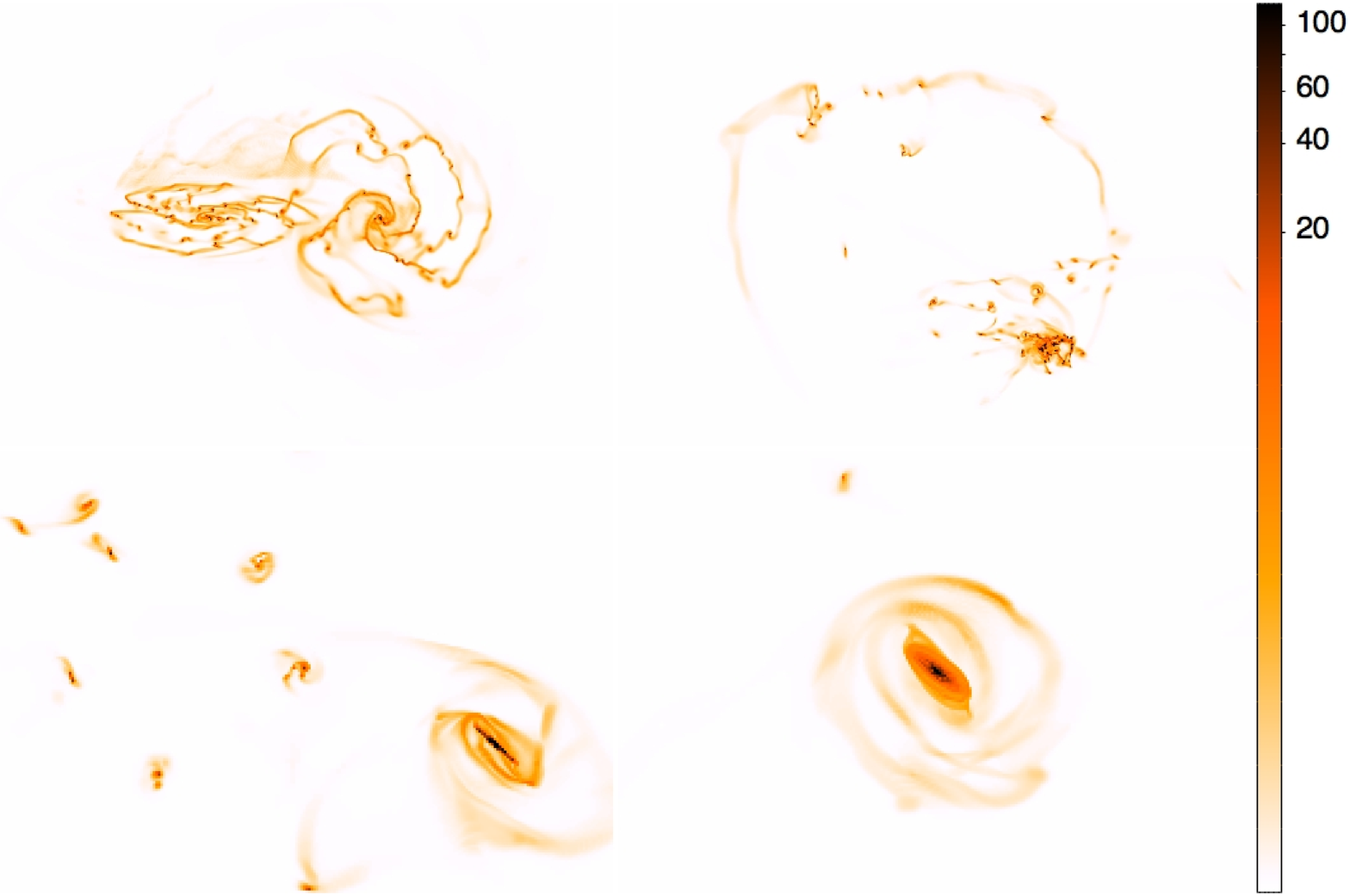}
\caption{ Time sequence of gas density maps showing the maximum gas density along the line of sight in units of H ${\rm cm^{-3}}$ for the mergers. For ease of comparison between the different mergers, the colour scale is limited at a density of $\sim 100$ H ${\rm cm^{-3}}$ in all images. Note that the actual maximum density in these maps is of this order of magnitude, which is considerably lower than the maximum density that can be reached overall. This is because these maps are extracted on level $11$ of the AMR grid, giving them an equivalent resolution of $\approx 80$pc (compared to the maximum possible $\approx 5$pc resolution on level 15). {\bf From top to bottom right:} Mergers A, B, C, D and E. }
\label{gasdens_timeseq}
\end{figure*}

\begin{figure*}
\includegraphics[width=\textwidth]{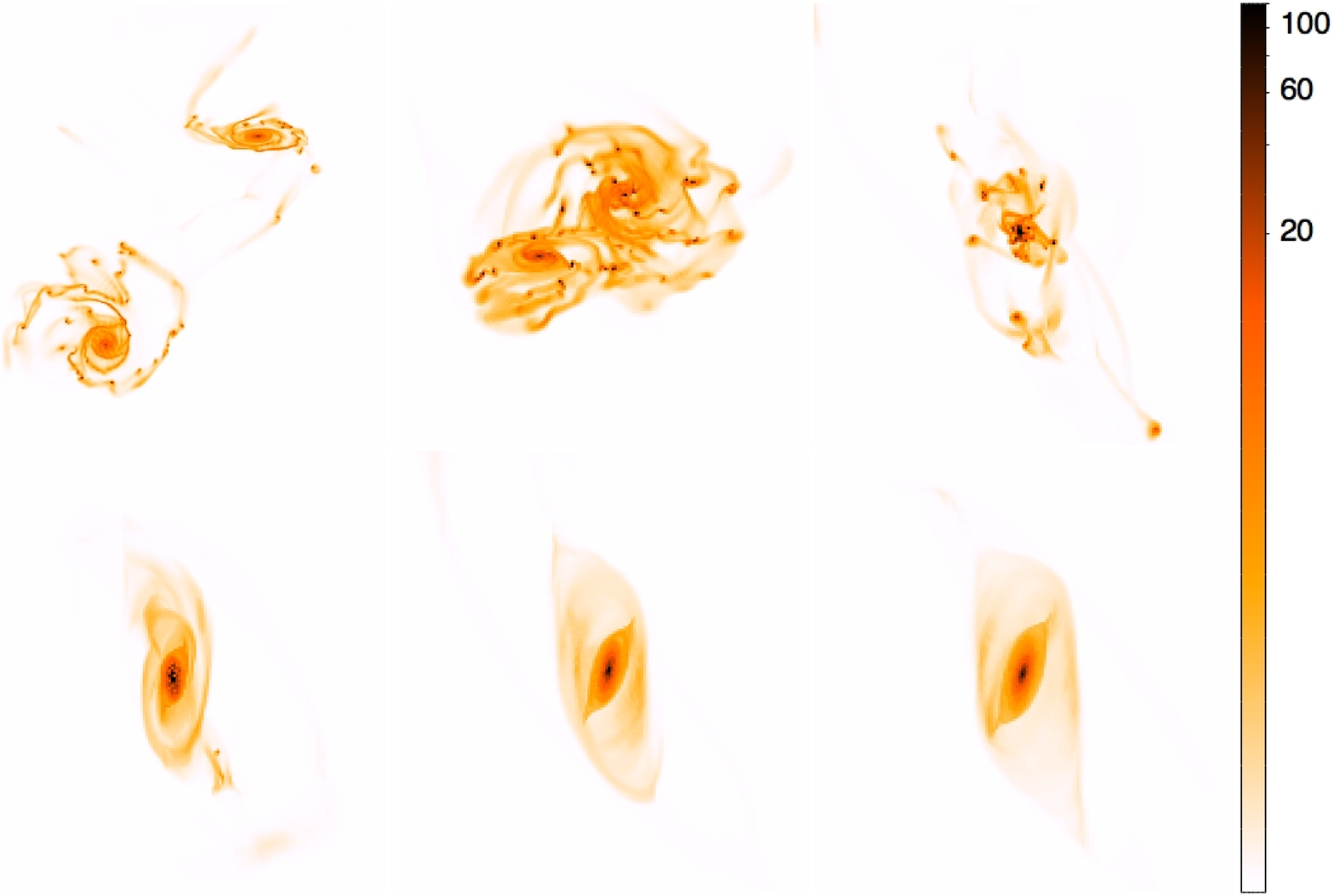}
\vskip 0.5cm
\includegraphics[width=\textwidth]{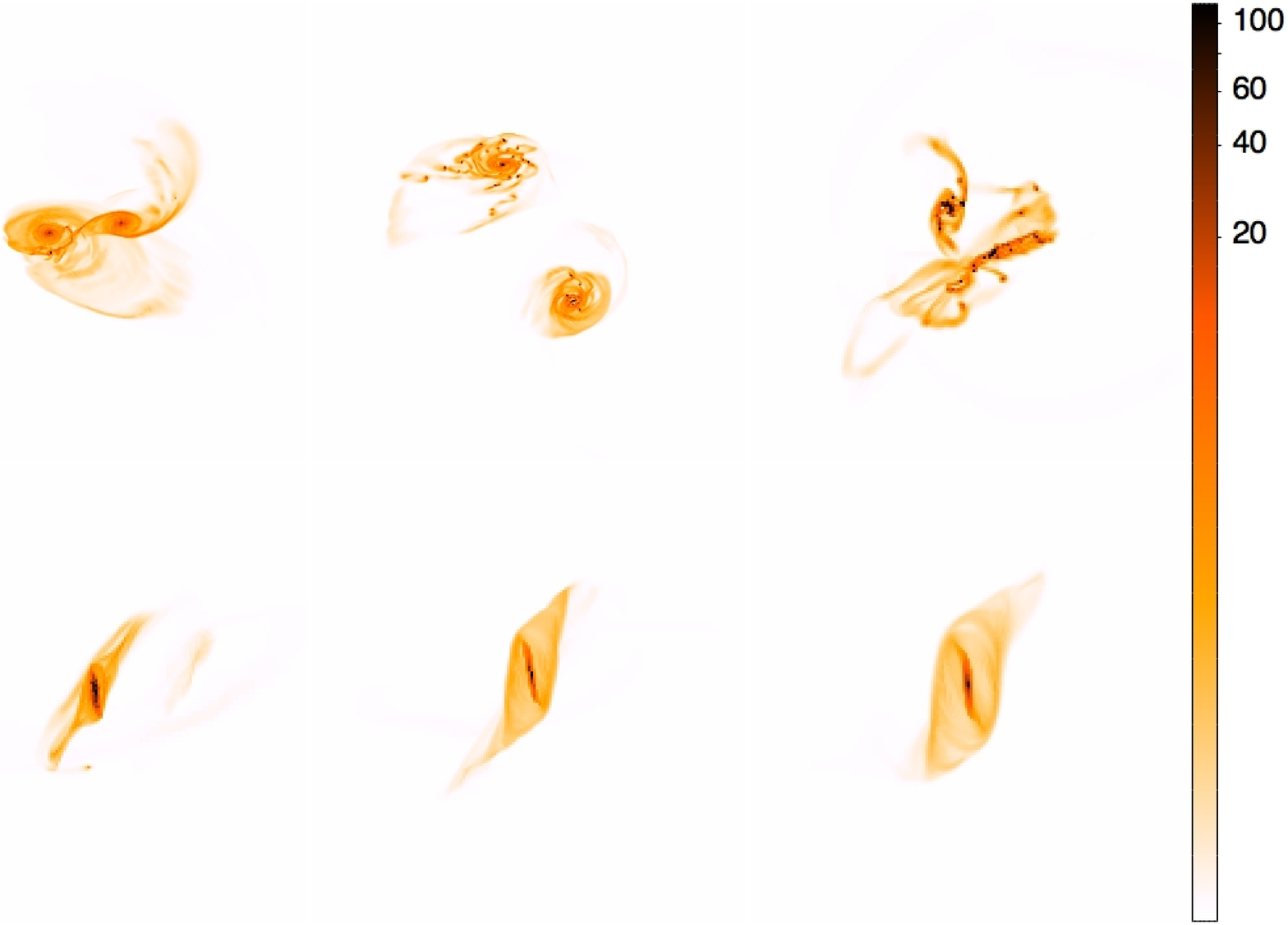}
\contcaption{}
\end{figure*}

\begin{figure*}
\includegraphics[width=\textwidth]{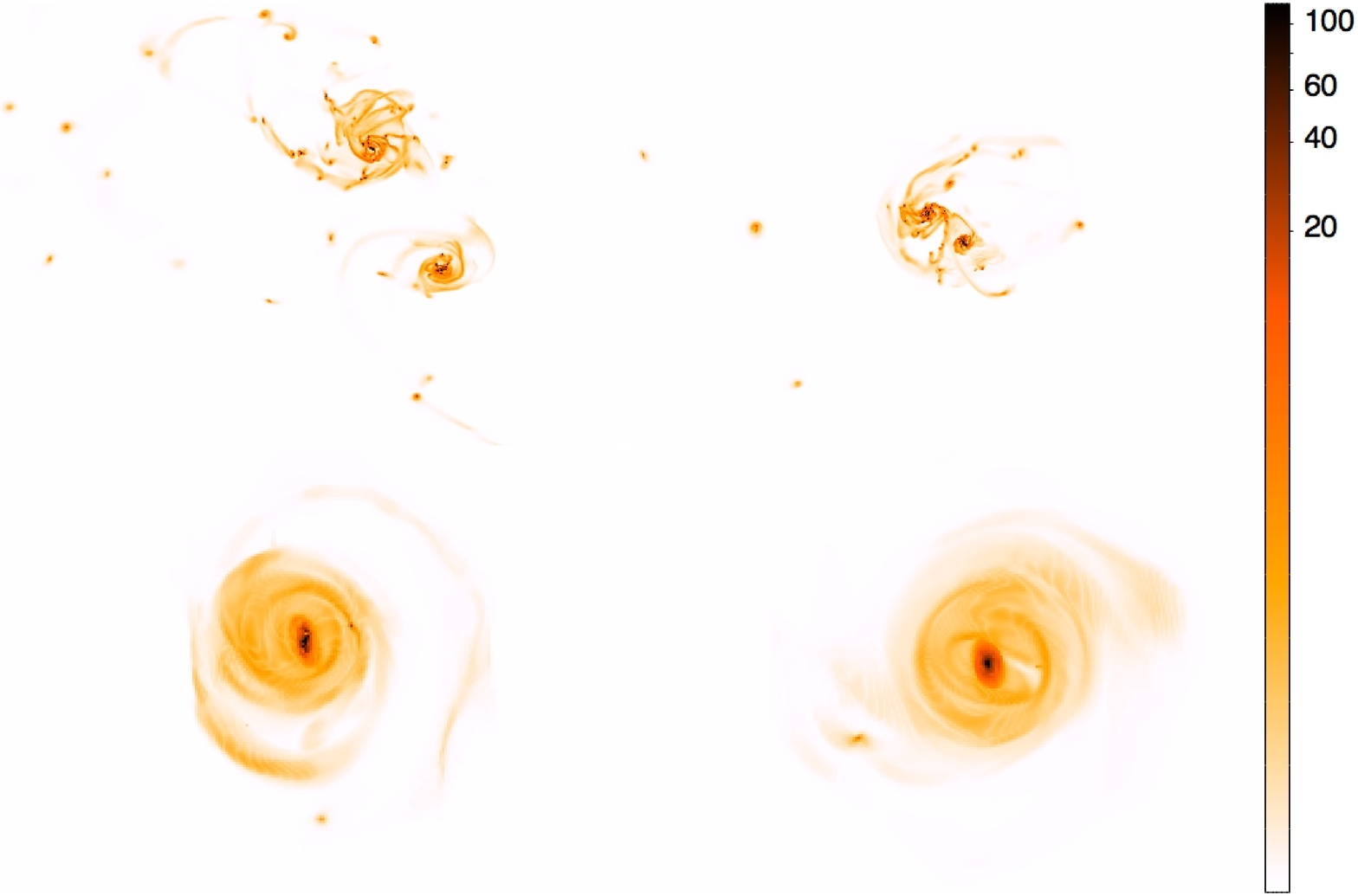}
\contcaption{}
\end{figure*}

\subsection{The simulations}

We perform the simulations with AMR code {\sc ramses} \citep{ramses}. We use a box-size of $160$kpc and a coarse grid of $64^3$ and allow up to $9$ further levels of AMR. This results in a maximum spatial resolution of $4.88$ pc in the densest regions. A grid cell is refined when there are more than 32 particles or the baryonic mass exceeds $1.28\times10 ^5 {\rm M}_{\odot}$. In the initial conditions, the dark matter particle mass is $1.2 \times10 ^5 {\rm M}_{\odot}$ and the star particle mass is $7.5 \times10 ^4 {\rm M}_{\odot}$. 

For the gas cooling we use an equation of state, which is discussed in detail in \citet{antennae_ramses} and \citet{bournaud_ism}. Essentially the equation of state gives the temperature for a solar metallicity gas of a given density when equilibrium is reached between atomic and fine structure cooling and UV heating from a UV background \citep{uvbackground}. In cells where the Jeans length is not resolved when our equation of state is applied, a polytropic equation of state is invoked instead to force this requirement, acting to prevent artificial fragmentation. Gas above a threshold density of $5 \times 10^3 {\rm cm}^{-3}$ forms stars according to a Schmidt law, $ \dot{\rho}_{\rm star} = \epsilon[\rho_{\rm gas}/t_{\rm ff}]$, where $t_{\rm ff}$ is the freefall time and $\epsilon$ is the efficiency, which we set to $0.2$ per cent. We include kinetic feedback using the supernova (SN) Sedov solution implemented in {\sc ramses} \citep[see][for details]{dubois_teyssier_sn}. In our simulations, the fraction of stellar mass recycled in each SN is $0.2$ and the initial blastwave radius is $10$ pc.

The initial conditions are set up identically for all galaxies, as follows. The dark matter sphere has a Plummer profile and is truncated at  a radius of $40$ kpc. The total gas  mass is $9 \times10^9 {\rm M}_{\odot}$ and the gas disc has an exponential profile, with a scale-length of $3$kpc (truncated at $15$kpc) and a scale-height of $100$pc (truncated at $900$pc). The gas fraction of the galaxies is initially $\approx 12$ per cent, but this decreases by the time the galaxies merge (due to gas consumption) and so is appropriate for simulating a low-redshift  ($z<1$) system. The gas fraction at the time of mergers varies between the different runs, so our study inherently explores a variety of combinations of orbits and gas fractions. The galaxies naturally develop SFRs of $1-5$ ${\rm M}_{\odot} {\rm yr}^{-1}$. We note that our merging galaxies do not have a reservoir of hot gas in the form of a hot halo, the impact of which was investigated recently in \citet{MosterMaccioSomerville2011}. We stress that the aim of this paper is to investigate the impact of the merger on the ISM structure, in the situation where its multiphase nature is properly captured, so we neglect possible sources of external gas accretion.

\begin{table}
\begin{tabular}[]{|c | c | c | c | c | c | c | c |}
\hline
Merger & {\it b} & $V_{\rm rel}$ & $E_{\rm orbital}=$ & {\it i}  & P/R  & {\it i} & P/R\\
 &(kpc)  &(km/s) & $\frac{E_k+E_g}{|E_k|+|E_g|}$&  G1 &G2  &  G1 &  G2 \\
\hline
A & 42 & 197 & -0.101 & 132&P&60&R\\
\hline
B  & 0 & 200 & -0.187&-&-&-&-\\
\hline
C & 42 & 197 & -0.101&60&R&83&R\\
\hline
D & 42 & 197 & -0.101& 47 &R & 120& P\\
\hline
E & 34& 145 & -0.423& 73& R & 120 & P\\
\hline
\end{tabular}
\caption{Table of orbital parameters for the merger sample. {\bf From left to right}: The impact parameter, $b$ (kpc), the relative velocity of the two galaxies, $V_{\rm rel}$ (km/s), the energy of the orbit, $E_{\rm orbital}$ (dimensionless), the angle of inclination of the disc to the orbital plane, $i$ (deg) and the type of orbit, prograde, P, or retrograde, R, for galaxies $1$ and $2$ (denoted by G$1$ and G$2$ respectively).}
\label{orbits}
\end{table}

Table \ref{orbits} gives the orbital parameters of all the mergers in our sample. The parameters were selected such that the galaxies are not aligned along any of the main axes (x,y,z), the mergers have large impact parameters and are mostly on nearly parabolic orbits (i.e. $E_{\rm orbital}$ is close to 0) such that they are the type of mergers that occur frequently in a $\Lambda$CDM cosmology \citep{KhochfarBurkert2006}. As a point of comparison, we also simulate a head-on collision in simulation B. Fig.~\ref{gasdens_timeseq} shows the time evolution of the gas density during the starburst for all the mergers (from top to bottom: A, B, C, D and E) to illustrate how the interactions progress during the period of greatest interest for this study.

\subsection{Identifying the galaxies}\label{identify}

To define and separate the galaxies, we use the `old stars', that is the stars from the initial conditions files. We know to which of the two galaxies each old star belongs based on its ID, so this provides a clean way to separately identify the two objects as the merger progresses. The centre for each galaxy is computed using an iterative procedure to find the centre of mass of the old stars. When the centres of galaxy $1$ and galaxy $2$ are less than $5$kpc apart, they are defined as a single object. Up to this point, gas and `new stars' (i.e. stars created during the simulation) are attributed to the galaxy whose centre they lie closest to.\\

\noindent In the following sections we focus primarily on the time periods during which the mergers are undergoing starbursts. All figures showing time evolution relate to the starbursts only (except Fig.~\ref{sfr}, which shows data for the duration of the simulations).

\section{Merger-induced star formation}

\subsection{Comparing global properties of star formation in isolated, pre- and mid- merger discs.}
\label{sf_compare}

\begin{figure}
\includegraphics[width=0.49\textwidth]{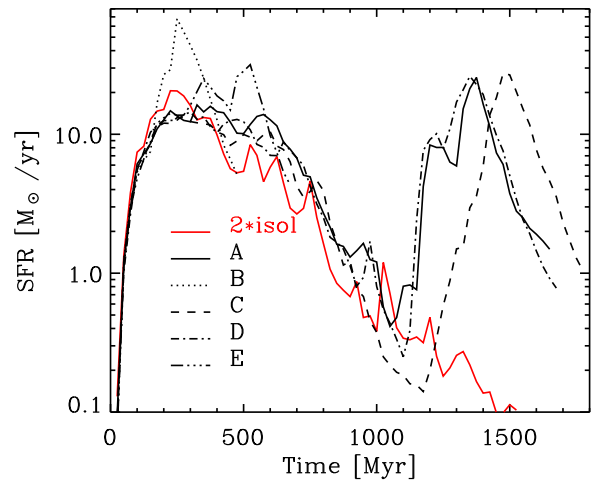}
\caption{SFRs for all the mergers and $2^*$SFR for the isolated galaxy (solid red line) for comparison. SFRs are for the whole box and are computed using $25$ Myr time bins. Mergers are A (solid black line), B (dotted line), C (dashed line), D (dash dot line) and E (dash triple-dot line.) } 
\label{sfr}
\end{figure}

One of the main points of interest when studying mergers is the high SFRs observed during the interaction i.e. starbursts. In Fig.~\ref{sfr} we compare the SFRs for all our merger simulations (SFRs are measured for the whole simulation box) compared to twice the SFR measured in the isolated galaxy (red line). We calculate the average value of the SFR in the isolated disc  to be $1.9{\rm M_{\odot}}$/yr over the time period $400$Myr to $1$Gyr (this period was chosen as it is when the SFR is in an appropriate range for a low redshift disc). This provides a baseline in order to determine the enhancement in the SFR during the mergers.

While earlier studies are able to resolve clumps during the merger, they are not able to do so in the pre-merger discs \citep[e.g.][]{antennae_ramses}. This is simply because the Jeans mass is small in the pre-merger discs (and therefore below their resolution limit), but it increases during the interaction. Due to our higher resolution, star formation is clustered in our isolated disc, confirming that we resolve the clumpy structure of the pre-merger discs before tidal torques etc come into play. We note that since the mergers occur after different amounts of time have elapsed, there is some variation in the ISM structure when the interaction takes place. As with the range of gas fractions discussed earlier, this variety in the pre-merger galaxy properties (i.e. more clumpy ISM and higher SFR, like an Sd, versus less clumpy ISM and lower SFR, like an Sa) is also advantageous, as it means our analysis is less tied to a specific type of pre-merger galaxy.

Two of the mergers, B and E, occur relatively quickly, within around $500$Myr of starting the simulation, whereas the other $3$, A, C and D occur at around $1500$Myr. In all cases a defined starburst (i.e. a sharp peak in the SFR) is clearly visible, indicating that the merging process has boosted the star formation for all the sets of orbital parameters. We typically measure enhancement factors, i.e. ${\rm SFR_{peak}}/{\rm 2<SFR_{isolated}>}$, of $\approx 10$ at the peak of the SFR (the actual values are $7$, $16$, $9$, $7$ and $10$ for mergers A, B, C, D and E respectively).  

\citet{Di-MatteoBournaudMartig2008} find that in two large sets of merger simulations (run with different computational techniques) only $5$\% of significant interactions or mergers result in an increase in star formation by a factor of $5$ or more. \citet{JogeeMillerPenner2009} find that the SFR is enhanced by only a factor of a few in interacting systems compared to isolated systems for $\sim 3600$ galaxies from {\sc gems}. We do not have a statistically significant  number of simulations, but can still confirm that our simulations would fit reasonably well with these larger studies. Considering that central star formation may be dust-obscured in observations of mergers and that older simulations do not have sufficient resolution to resolve clumpy star formation, our results are not in conflict with previous findings drawn from large samples of merging galaxies.

\subsection{The gas response to the merger}

\subsubsection{Density}

\begin{figure*}
\includegraphics[width=0.33\textwidth,trim = 10mm 3mm 10mm 5mm,clip]{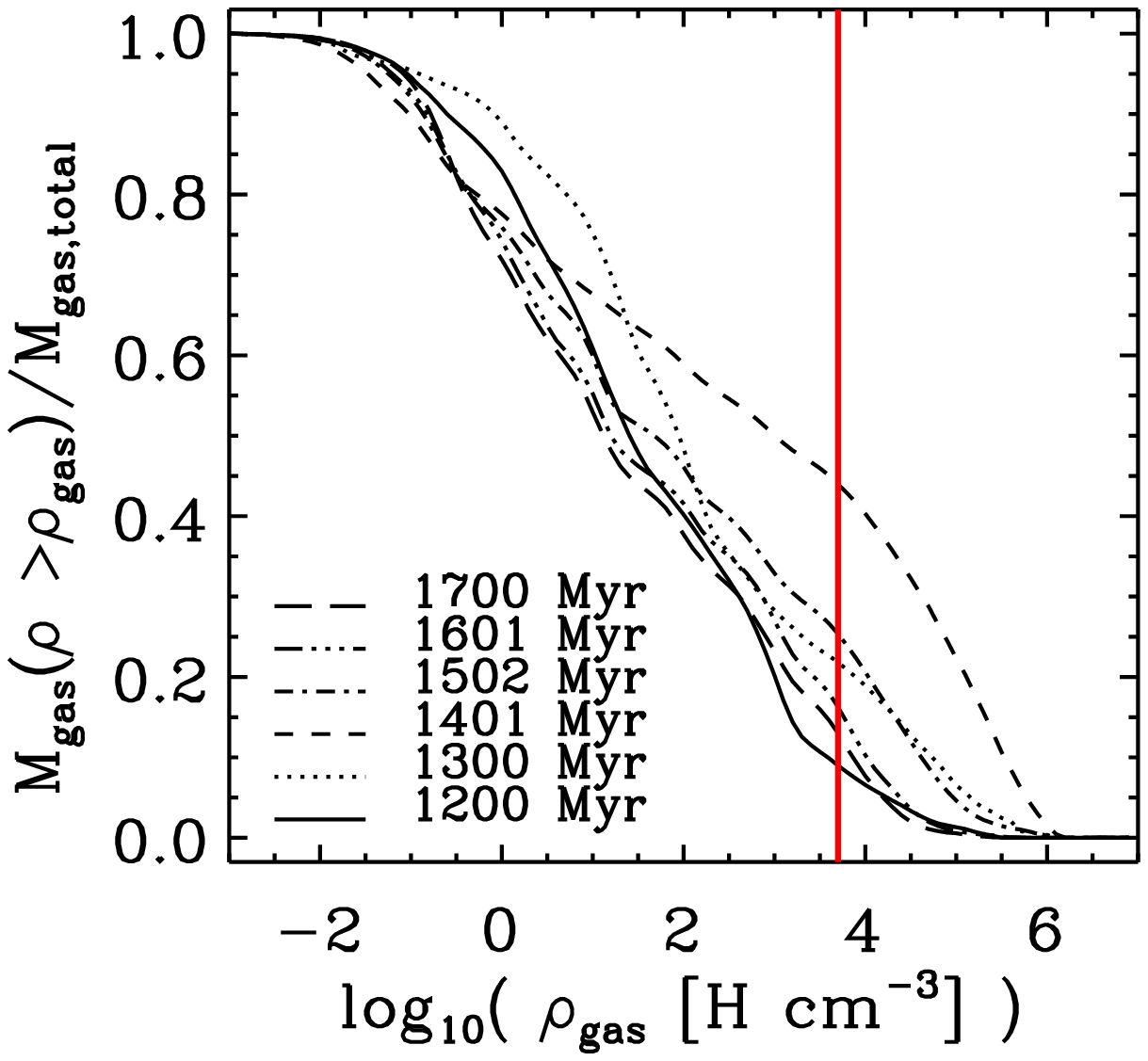}
\includegraphics[width=0.33\textwidth,trim = 10mm 3mm 10mm 5mm,clip]{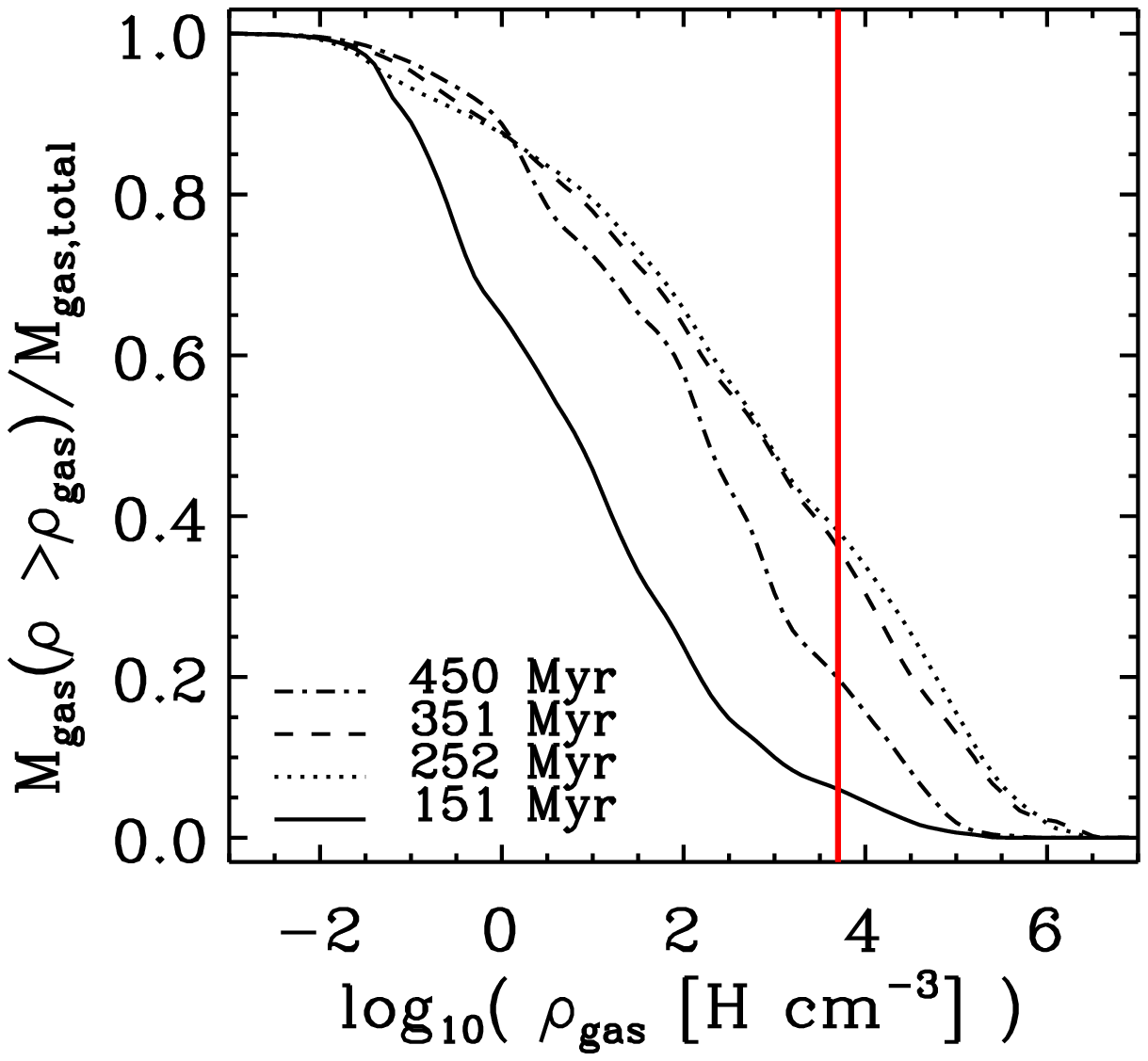}
\includegraphics[width=0.33\textwidth,trim = 10mm 3mm 10mm 5mm,clip]{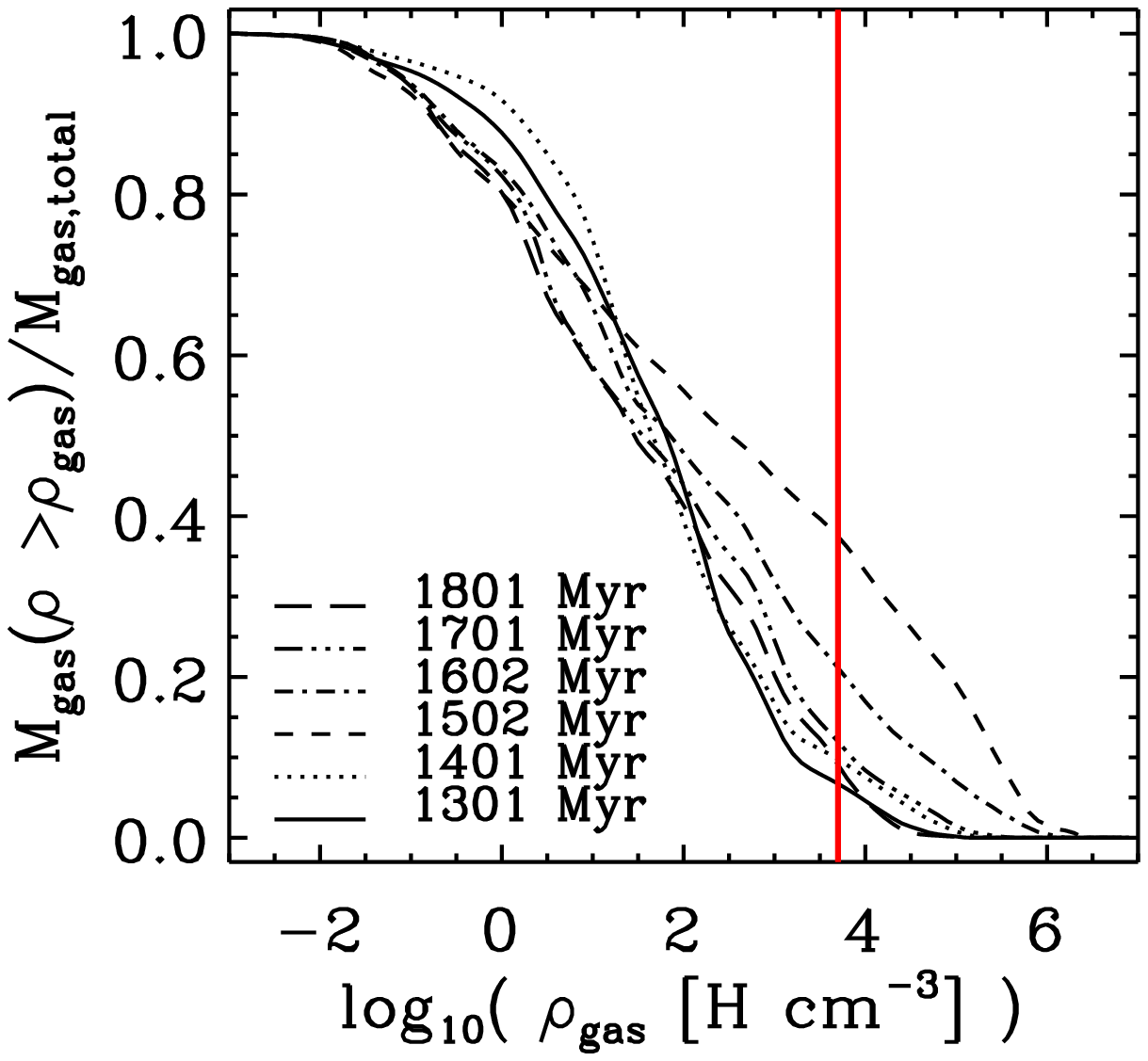}\\
\includegraphics[width=0.33\textwidth,trim = 10mm 3mm 10mm 5mm,clip]{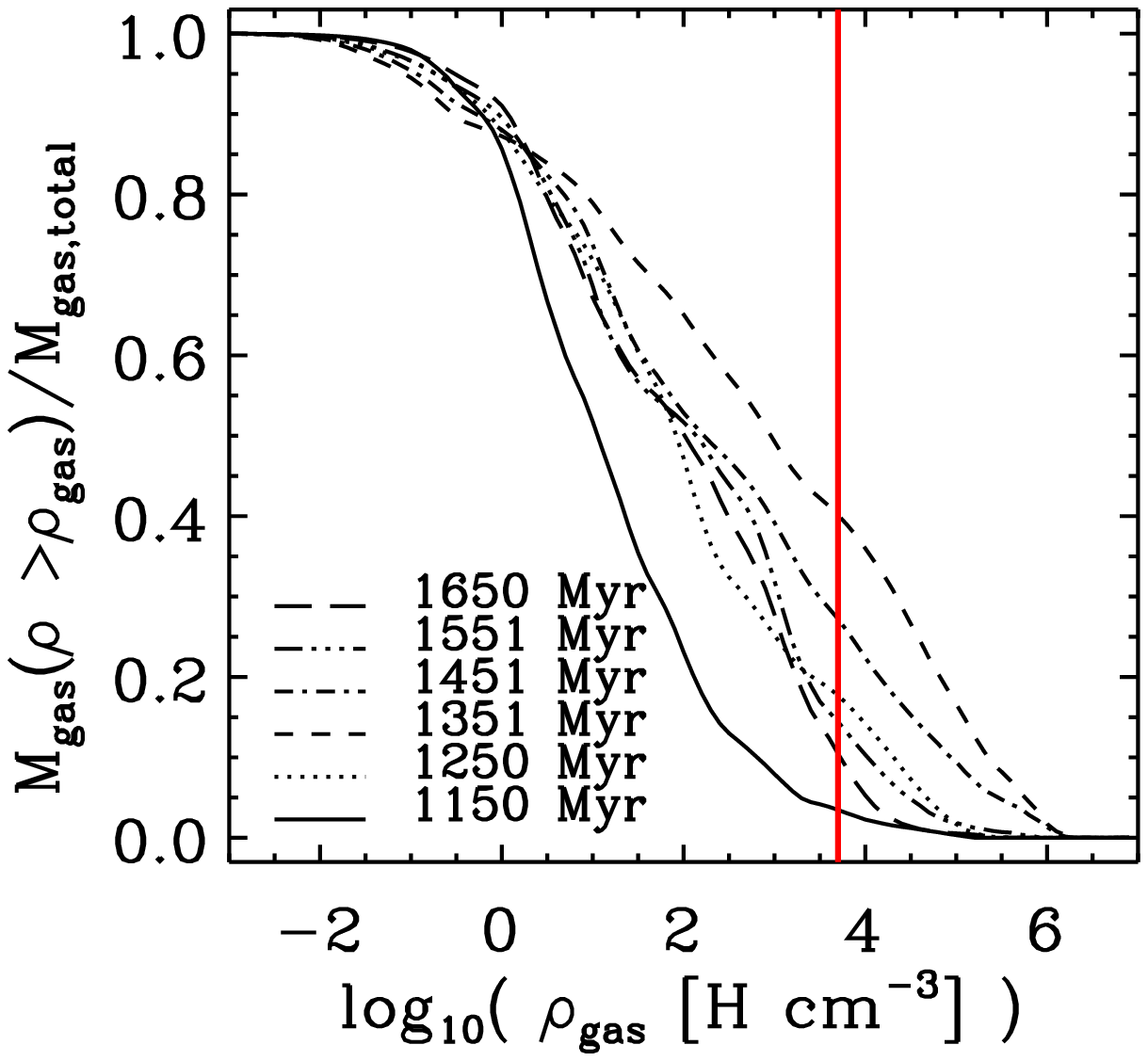}
\includegraphics[width=0.33\textwidth,trim = 10mm 3mm 10mm 5mm,clip]{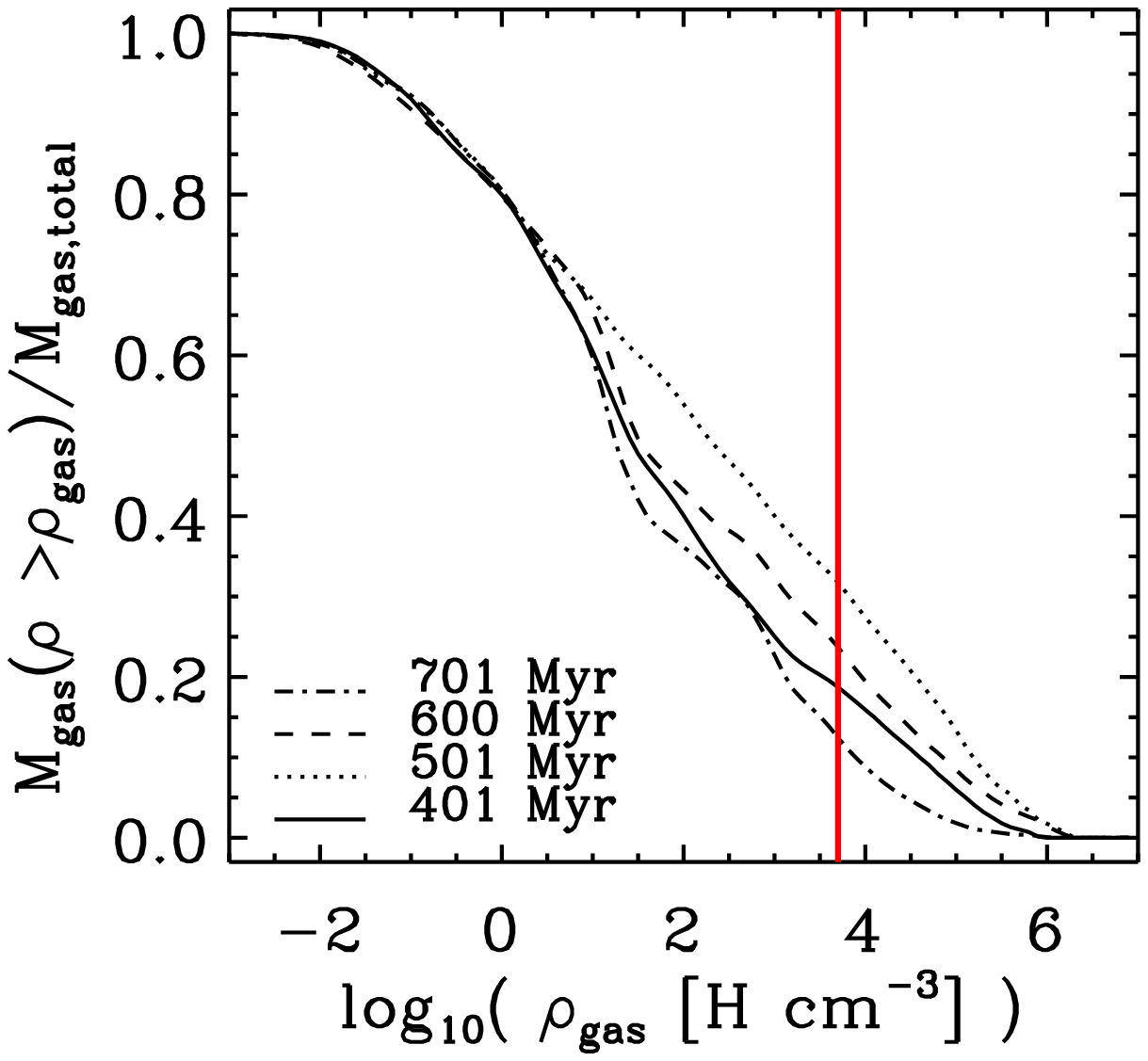}
\caption{Density PDF of the gas within a $15$kpc radius in gal$1$, for each of the mergers. The vertical red lines show the density threshold for star formation used in the simulations. {\bf From top left to bottom right:} Mergers A, B, C, D and E. }
\label{pdf_all}
\end{figure*}

Knowing how the gas density in merging galaxies changes during the interaction is crucial to our understanding of the star formation mechanisms, since at the most basic level, dense gas becomes stars. Fig.~\ref{pdf_all} shows the time evolution of the gas density for galaxy $1$ in all the mergers during their starbursts (from top left to bottom right: A, B, C, D and E). The gas density probability density function (PDF) undergoes strong evolution in all the mergers, with the appearance of a significant excess of gas at high densities ($\rho > 10^4$ H ${\rm cm}^{-3}$). An excess of dense gas as the merger evolves is also seen in simulations of the antennae system \citep{antennae_ramses}. We stress that the simulations in the current work now include SN feedback (in contrast to the Antennae simulations) and so the density excess is a robust result.

The specific way the density PDFs evolve in our simulations can explain observations of enhanced HCN/CO ratios in ULIRGs without AGN, one of the reasons often provided for the enhancement \citep[e.g.][]{Gracia-CarpioGarcia-BurilloPlanesas2008}. The typical densities traced by CO emission are $n$ $\gtsim $ $300$ ${\rm cm}^{-3}$ whereas HCN traces densities two orders of magnitude higher, at $n$ $\gtsim$ $ 3\times10^4$ ${\rm cm}^{-3}$ (\citet{GaoSolomon2004}, see also \citet{JuneauNarayananMoustakas2009} and references therein for further discussion). It is clear, that based on the behaviour of the density alone (without making detailed radiative transfer calculations of the emission), we would expect enhanced HCN/CO ratios. \citet{JuneauNarayananMoustakas2009} used hydrodynamical simulations  combined with radiative transfer, to demonstrate that HCN/CO ratios could be enhanced in mergers (relative to isolated discs) due to increased dense gas fractions. They did not resolve the multiphase ISM however, using a subgrid model for Giant Molecular Cloud formation instead. It is encouraging that we find a significant increase in the dense gas fraction in our simulations, in which the dense clouds are simulated directly.

\begin{figure}
\includegraphics[width=0.49\textwidth]{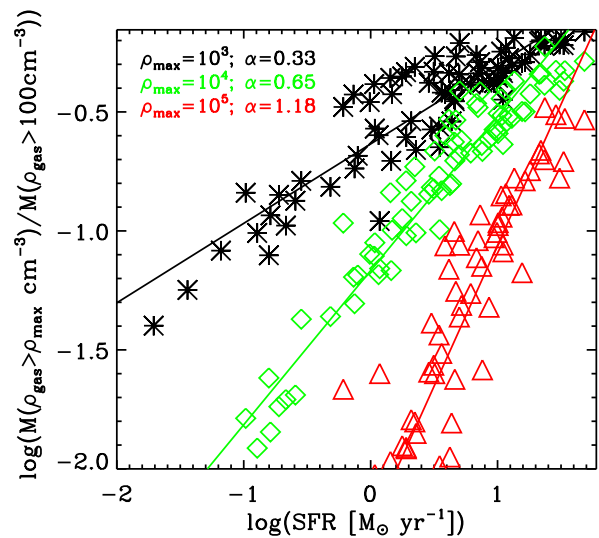}
\caption{ Mass ratio of gas with $\frac{\rho>\rho_{max}} {\rho>100 {\rm H cm^{-3}}}$ for $\rho_{max}= 10^3 $ H $ {\rm cm^{-3}}$ (black asterisks), $10^4$ H $ {\rm cm}^{-3}$ (green diamonds) and $10^5$ H $  {\rm cm}^{-3}$ (red triangles) versus SFR, for all mergers combined. The best-fitting line for each density ratio is also shown and the gradient, $\alpha$, for each is given in the legend.} 
\label{hcn2co}
\end{figure}

To examine this issue in more detail, we plot the ratio of  the mass of gas with density greater than $10^3{\rm cm}^{-3}$ (black asterisks), $10^5{\rm cm}^{-3}$ (green diamonds) and $10^5{\rm cm}^{-3}$ (red triangles) to the mass of gas above the critical density of CO ($\sim 100{\rm cm}^{-3}$) versus SFR in Fig.~\ref{hcn2co}. The data for both galaxies, in all of the mergers, (sampled at equally spaced time intervals during the starbursts) has been combined. We note that in the case of the highest density threshold, $\rho_{max}=10^5{\rm cm}^{-3}$, not all of the outputs have gas at these densities and so these points have been excluded from the plot, leaving a slightly smaller sample for this measurement. 

In many hydrodynamical simulations, including those presented here, the SFR is by definition linked to the mass of dense gas, since a minimum gas density criterion for star formation is applied. It is still interesting, however, that we find such a tight correlation for all the values of $\rho_{max}$. The star formation density threshold in the simulations is  $5 \times 10^3 {\rm cm}^{-3}$, so the  $\rho_{max}$ values probe gas both above and below this, yet the scatter is similarly small in all cases. The mass ratios are good proxies for the luminosity ratios of various dense gas tracers and the SFR is a reasonable proxy for the infrared luminosity, allowing us to compare with observations.  \citet{JuneauNarayananMoustakas2009} find best-fit slopes, $\alpha$, ranging from $\approx 0.23-0.69$ and our values of $\alpha=0.33,0.65,1.18$ are compatible with this (compare their Fig.~6 with our Fig.~\ref{hcn2co}). We also see the same significant trend of an increasing best-fit slope, $\alpha$, with increasing density ratio, as shown in Fig.~7 of  \citet{JuneauNarayananMoustakas2009}. Our simulation work therefore supports their hypothesis that the density distribution of gas determines molecular line ratios and shows that the evolution of the density PDF during mergers could explain enhanced luminosity ratios in ULIRGs.

It is evident from the time sequences of gas density maps in Fig.~\ref{gasdens_timeseq} that the distribution of the densest gas varies significantly between the mergers  and indeed in any given merger during the course of its starburst. This suggests that studying the gas response and star formation at the peak of the starburst alone is not necessarily representative of the mechanisms at work throughout the starburst. We examine the distribution of star formation in more detail in Section \ref{morph}.

\subsubsection{Velocity dispersion}

\begin{figure*}
\centering
\includegraphics[width=0.32\textwidth,trim = 7mm 10mm 5mm 10mm,clip]{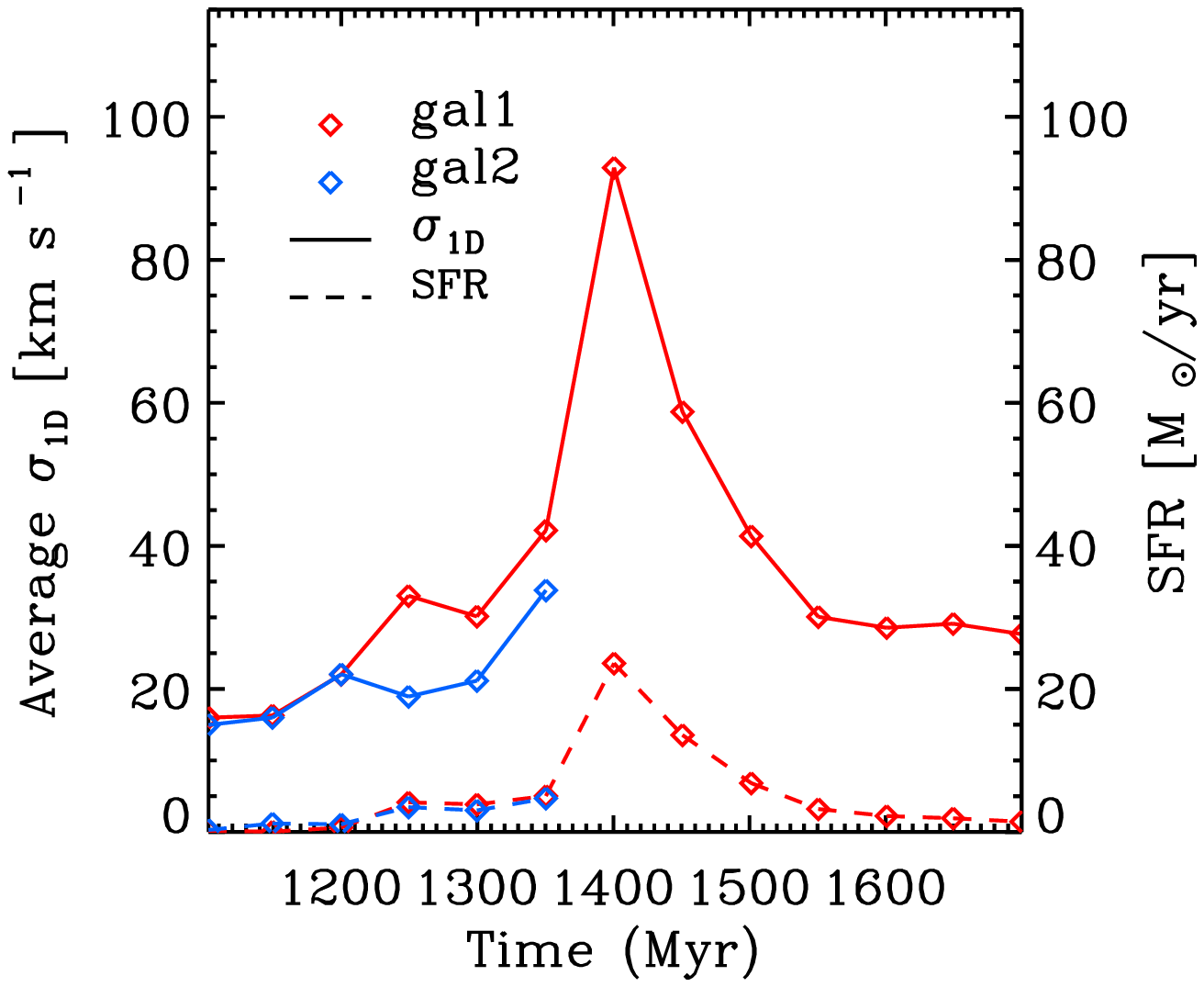}
\includegraphics[width=0.32\textwidth,trim = 7mm 10mm 5mm 10mm,clip]{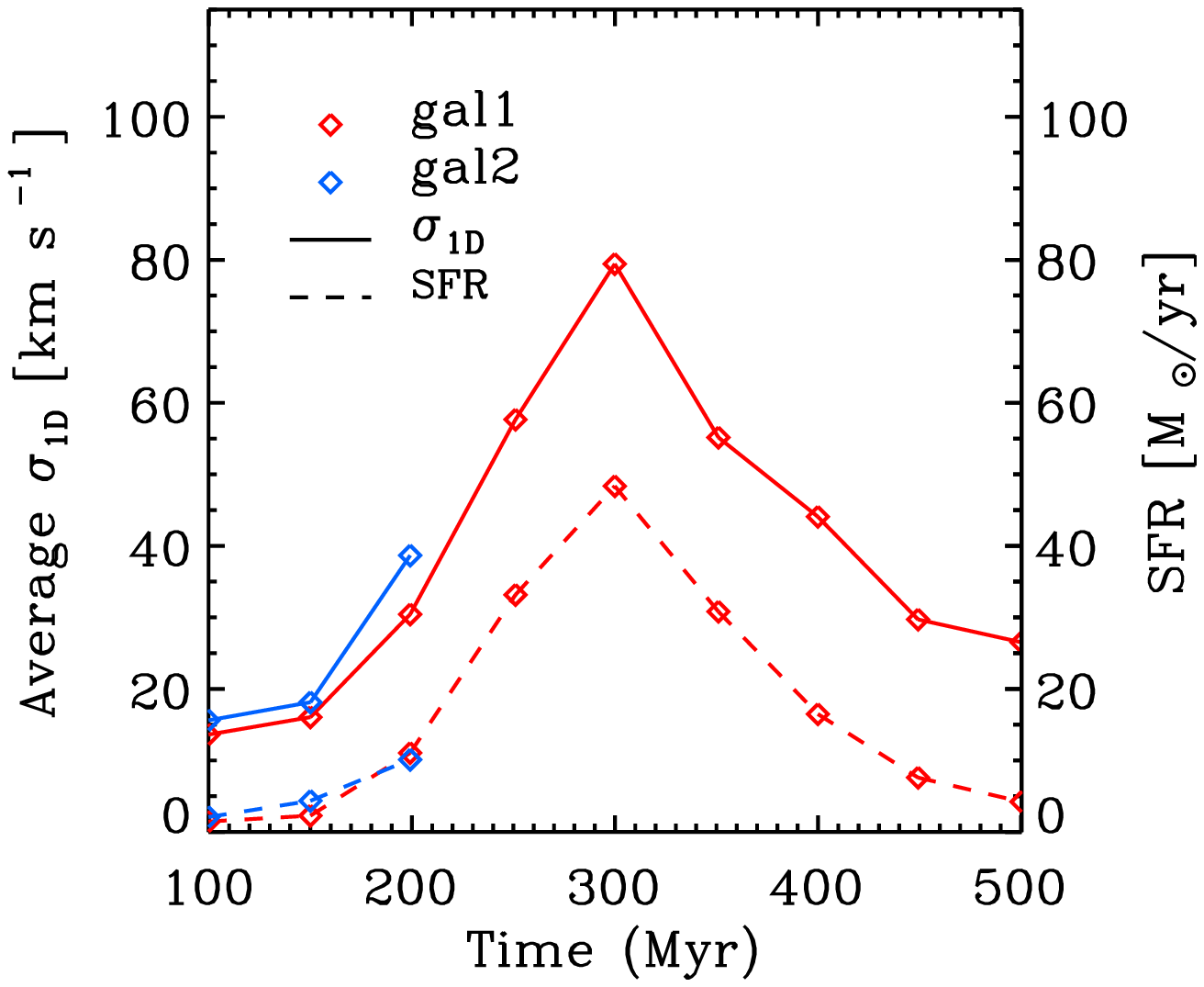}
\includegraphics[width=0.32\textwidth,trim = 7mm 10mm 5mm 10mm,clip]{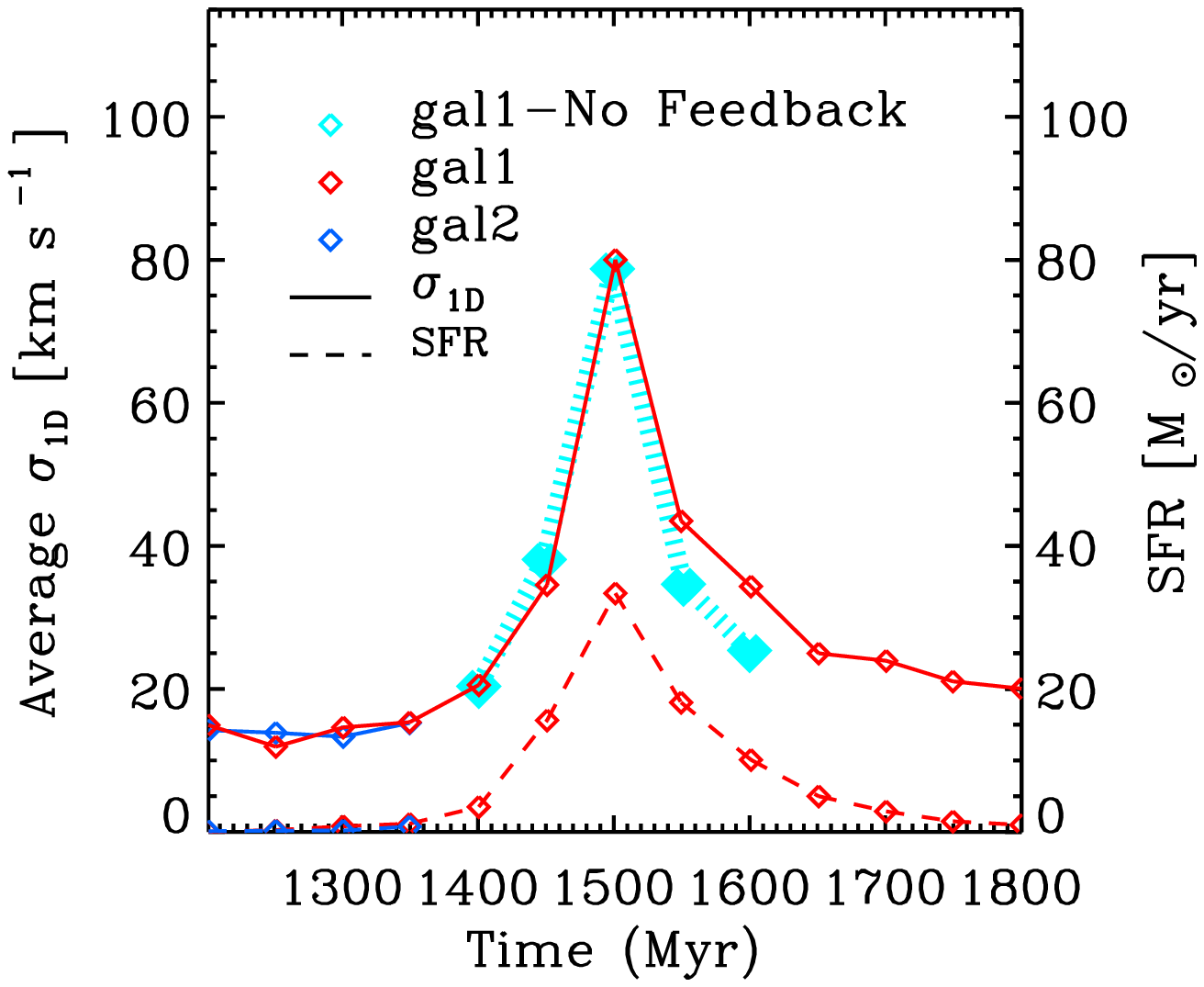}\\
\includegraphics[width=0.32\textwidth,trim = 7mm 10mm 5mm 10mm,clip]{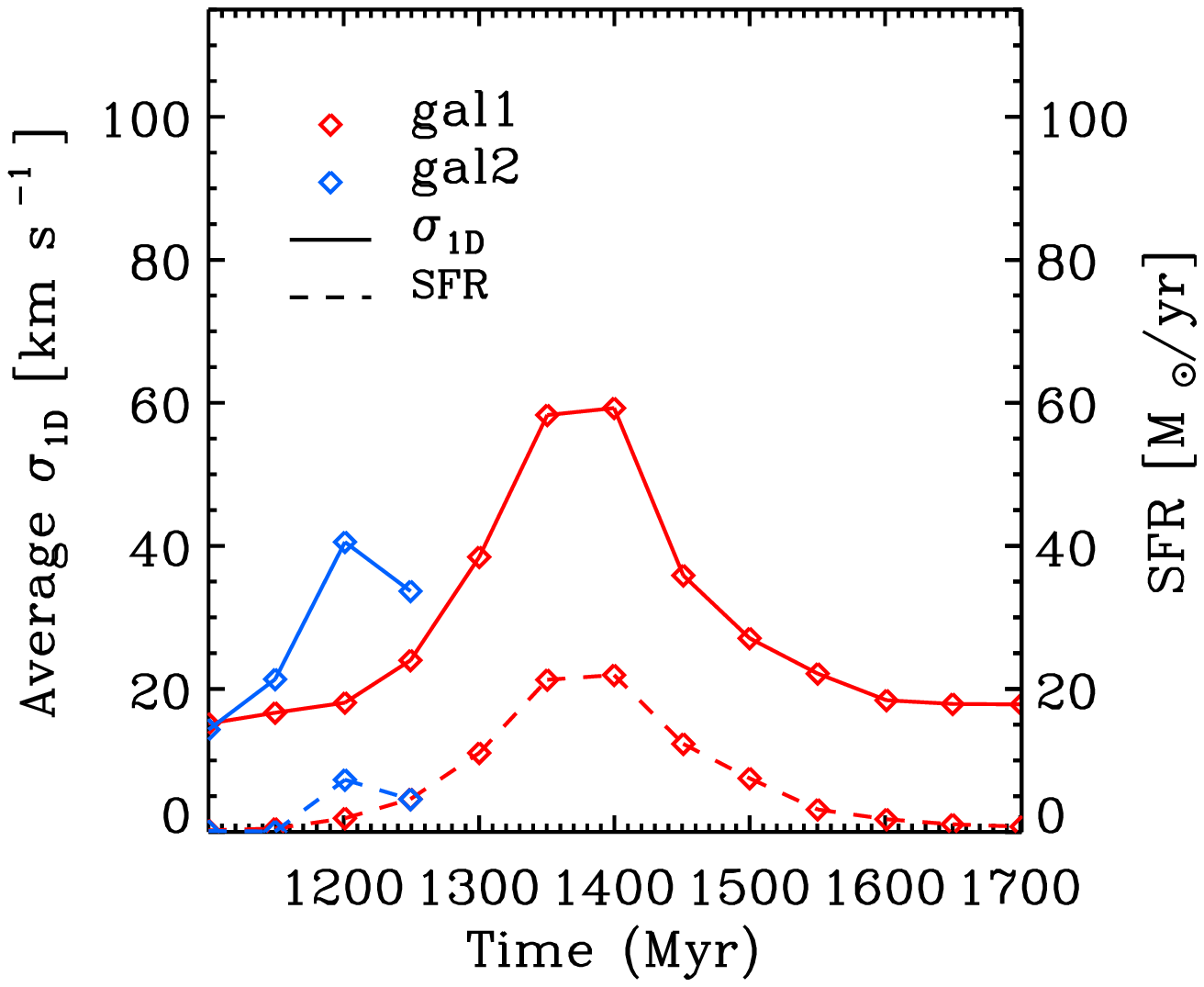}
\includegraphics[width=0.32\textwidth,trim = 7mm 10mm 5mm 10mm,clip]{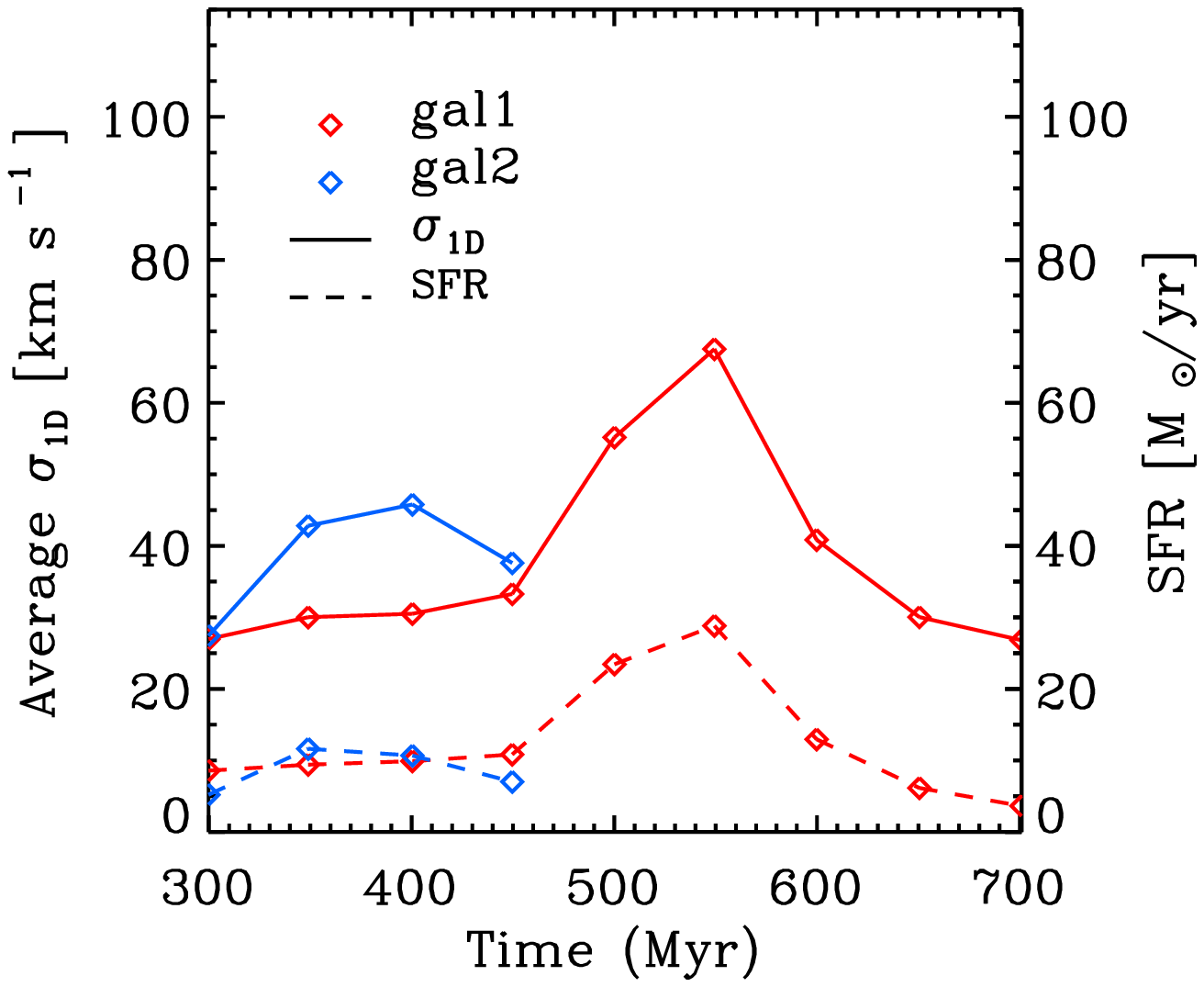}

\caption{Time evolution of the average velocity dispersion (solid lines) and SFR (dashed lines) for both galaxies (galaxy $1$ in red, galaxy $2$ in blue).  Measurements are taken within a $15$kpc radius around each galaxy centre.  {\bf From top left to bottom right:} Mergers A, B, C, D and E.  The additional  thick, broken, cyan line for merger C (top right panel) shows the average velocity dispersion when the simulation is rerun for the duration of the starburst with SN feedback switched off. This indicates that the interaction itself, rather than SN feedback, is driving turbulence in the gas. See text for further details.}
\label{veldisp}
\end{figure*}

Another important gas property to assess is the velocity dispersion, as this is observed to be higher in interacting systems. We measure the velocity dispersion for each galaxy as follows. A cube of side length $30$kpc is placed on the centre of the galaxy and is further divided into subcubes of side $100$pc. For every one of these subcubes that contains 10 or more AMR grid cells we compute a velocity dispersion, $\sigma_{\rm 1D}$, where $\sigma_{\rm 1D}=\mod{\sigma_{\rm 3D}}/\sqrt{3}$. For the purpose of the calculation the AMR cells are treated as pseudo gas particles i.e. the properties of each cell (mass, velocity etc.) are assigned to the coordinates of its centre. We derive a galactic value of $\sigma_{\rm 1D}$ by taking the mass-weighted average of the $\sigma_{\rm 1D}$ values computed for each subcube. The evolution of $\sigma_{\rm 1D}$ with time for galaxy $1$ (solid red lines) and galaxy $2$ (solid blue lines) of each merger is shown in Fig.~\ref{veldisp}. During all of the mergers $\sigma_{\rm 1D}$ increases dramatically, from around $20$km/s (a value maintained consistently in the isolated galaxy) to $60-80$km/s. 

Observations of interacting galaxies also exhibit velocity dispersions that are higher than in non-interacting galaxies, for which typical values of  $10$ km ${\rm s}^{-1}$ are measured \citep[e.g.][]{TamburroRixLeroy2009}. \citet{Irwin1994} measures an average velocity dispersion of $\approx 20$ km/s in both the interacting galaxies NGC5775 (a starburst galaxy) and NGC5774 (a barred spiral), but some regions reach a velocity dispersion of up to $50$km/s in the latter. In observations of the interacting galaxy NGC2207, \citet{ElmegreenKaufmanBrinks1995} measure velocity dispersions of $40-50 {\rm km s}^{-1}$ over a large area of the disc and also observe several large HI cloud complexes in the same region. They propose these have formed via gravitational instabilities as outlined in \citet{ElmegreenKaufmanThomasson1993}.  We note that in both cases the galaxy pairs are only at the initial stages of their interaction (before the SFR peaks and, therefore, probably before the velocity dispersion peaks) and so our maximum velocity dispersion measurements of $\approx 80 {\rm km s^{-1}}$ are consistent with these measurements. 

\subsubsection{Origin of increased velocity dispersion.}
\label{origin_vdisp}

There are two likely mechanisms for increasing the velocity dispersion in our merger simulations; the interaction itself or the `stirring up' of the ISM caused by SN explosions. In Fig.~\ref{veldisp} we also show the SFR (dotted lines) for the same region in which the velocity dispersions (solid lines) are measured. Since the peak in both the velocity dispersion and the SFR occur almost simultaneously in all the mergers, it is impossible to distinguish whether the interaction increased the turbulence, resulting in more star formation, or whether the star formation increased for another reason (e.g. global inflow and compression) and the resulting SN caused the increased turbulence. 

In order to disentangle these two possibilities we rerun the simulation for C (chosen at random) for the period of the starburst ($\sim1400-1600$Myr) during which time we halt SN feedback, although note that there was feedback prior to this period. This allows us to make the most accurate determination of the impact of the feedback on the turbulence. If we reran the whole simulation without feedback, both the gas fraction and structure of the ISM would be different prior to the starburst, making the comparison much more complex. We have chosen the time interval for the rerun to be long enough such that any boost in velocity dispersion after this interval must be from another source; if the feedback was driving the turbulence, we would expect the turbulence to decay $\sim 10$Myr after the feedback is switched off \citep{Mac-LowKlessenBurkert1998}.

In Fig.~\ref{veldisp} (top right panel) we show the velocity dispersion in galaxy $1$ for the original run of merger C (solid red line) and the corresponding velocity dispersion for the same galaxy in the `no feedback'  rerun (solid cyan line). The velocity dispersion measurement is almost identical to that when SN feedback was switched on, yet now the only possible driver of the significantly increased turbulence in our simulation is the interaction during the merger itself. \citet{HerreraBoulangerNesvadba2011} also deduce from observations of the Antennae system that it is the interaction, rather than SN explosions, that are driving turbulence in the ${\rm H}_2$ gas. Note that the increase in the SFR is still correlated with the increase in turbulence, suggesting the former may be being driven by the latter. 

With our high-resolution hydrodynamical simulations we are limited to studying the relationship between the velocity dispersion and the SFR in only a few galaxies. Results from recent semi-analytical models (SAMs), however, can provide insight into how this dependency would affect the whole galaxy population on cosmological timescales.  \citet{KhochfarSilk2009} show that setting the star formation efficiency (SFE) proportional to the gas velocity dispersion in their SAM results in better agreement with the normalisation of the observed ${\rm M}_{*}-$SFR relation at $z\sim2$. In their model, cold accretion (rather than mergers) is assumed to drive turbulence, resulting in higher velocity dispersions, higher SFEs and more star formation. Since SAMs cannot model the small-scale behaviour of the ISM, we are unable to say whether turbulence driven by external accretion would result in the same gas properties, such as the behaviour of the density PDF, that we have shown in the case of merger-driven turbulence. The mechanisms at work, therefore, may not be identical in these two scenarios. What we can conclude, however, is that the physical link we have demonstrated between the velocity dispersion and the SFR  can give rise to the observed cosmological star formation histories of the galaxy population.

\subsubsection{The relationship between the gas density PDF and turbulence.}

In order to gain some insight into possible links between the evolution of the gas density and the evolution of the velocity dispersion (properties examined in previous sections), we briefly review the current understanding of the impact of turbulence on the properties of the ISM \citep[for detailed reviews, see][]{elmegreen_scalo_2004,maclow_klessen_2004}.

It has been demonstrated many times using 2D hydrodynamical simulations that the density of isothermal gas subject to turbulence will approximately follow a lognormal PDF \citep[e.g.][]{vazquezsemadeni_1994}. A direct correlation between the width of the lognormal for isothermal gas, $\sigma$, and the Mach number of the supersonic turbulence, M is exhibited in simulations; $\sigma^2\approx\ln(1+3M^2/4)$ \citep[][and references therein]{KrumholzThompson2007}. The higher the Mach number of the turbulence, the higher mass fraction of gas at the highest and lowest densities i.e. a complex structure of over- and under- dense regions develops. The first simulations neglected self-gravity, thereby demonstrating that turbulence alone can cause fragmentation. 

Improving on earlier work, \cite{wada_norman_2001} find that in 2D simulations of the central region of a disk galaxy (which include self-gravity, heating and cooling processes and star formation) the ISM exhibits a perfect lognormal over 7 orders of magnitude in density above the mean. More recently it has been shown that the ISM density can be fitted by a lognormal PDF over several orders of magnitude in full 3D galaxy simulations \citet{TaskerBryan2008,WadaNorman2007}. These results suggest that the ISM in real galaxies may also show the same correlation between density PDF and turbulence as found in the very idealised ISM simulations.

\citet{FederrathRoman-DuvalKlessen2010} undertake a detailed study of isothermal supersonic turbulence in the ISM in simulations (examining both solenoidal and compressive forcing) and compare the results with observations. Their results present a more complex picture than the earlier work discussed above. They find that the properties of the turbulent gas differ considerably in different regions of the ISM subject to different combinations of  the types of forcing. In particular, while there is still a correlation between the width of the density PDF, $\sigma$, and the Mach number of the gas, $\sigma$ is around $3$ times larger for compressive forcing. Both types of turbulent forcing also result in PDFs that are only {\it approximately} lognormal. The authors also stress that the deviation from a lognormal density PDF is expected to be even more pronounced if self-gravity \citep[e.g.][]{Klessen2000} and non-isothermality \citep[e.g.][]{PassotVazquez-Semadeni1998} are taken into account (both relevant in the simulations presented in this work).

In our simulations, the density PDFs are, initially, approximately lognormal. However, due to the radial density gradient in the disk and features such as spiral arms, the mean density is different at different locations so average PDFs of the whole disk exhibit multiple peaks. Despite this, the PDFs still loosely follow the pattern described above; as the turbulence increases the PDF gets wider with more gas at the extreme ends of the density range i.e. very low density and very high density. In our case, however, the PDF shape also changes, with a second peak arising at high densities. This could possibly be due to a change in the type of turbulent forcing driven by the interaction \citep[e.g. see Fig.~4][]{FederrathRoman-DuvalKlessen2010} or a reflection of local collapse \citep[see Fig.~8(a)][]{Klessen2000}. As the turbulence increases during the merger, the mass fraction of dense gas increases and the SFR is driven up (see Fig.~\ref{veldisp}). It is possible, then, that what we are seeing is a type of turbulent fragmentation process.

\subsection{Distribution of the dense gas and the star formation}
\label{morph}

\begin{figure*}
\includegraphics[width=0.32\textwidth]{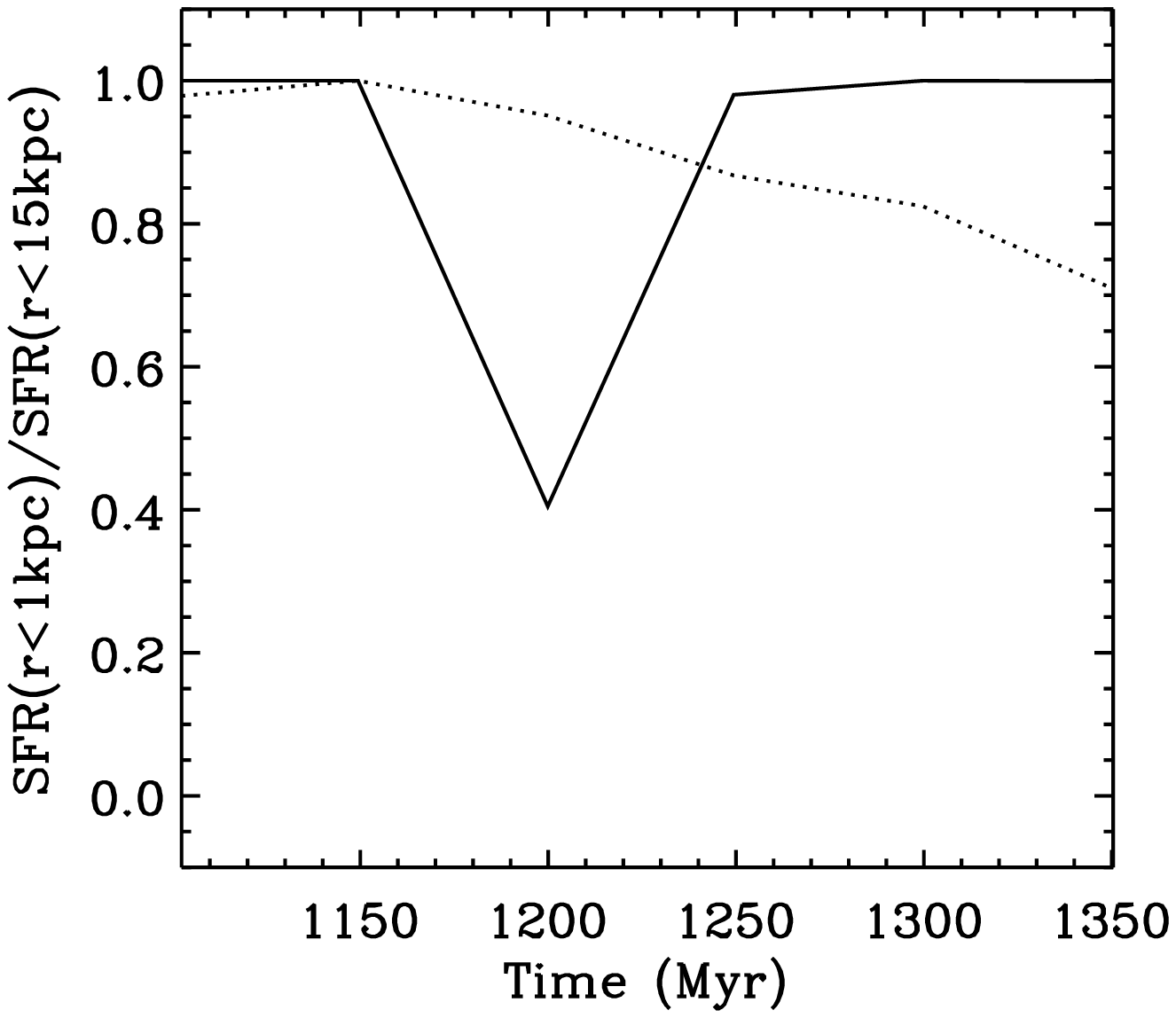}
\includegraphics[width=0.32\textwidth]{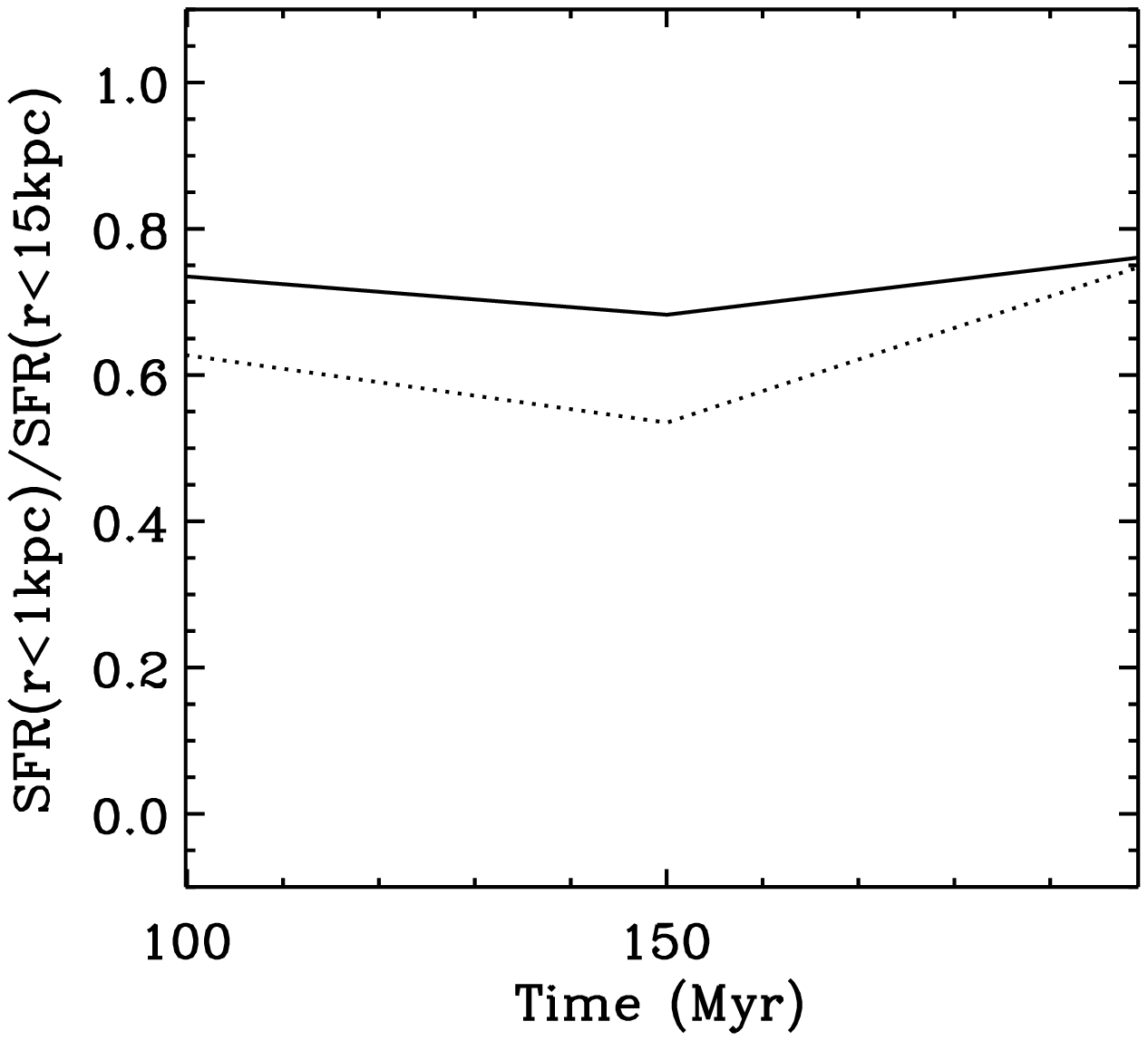}
\includegraphics[width=0.32\textwidth]{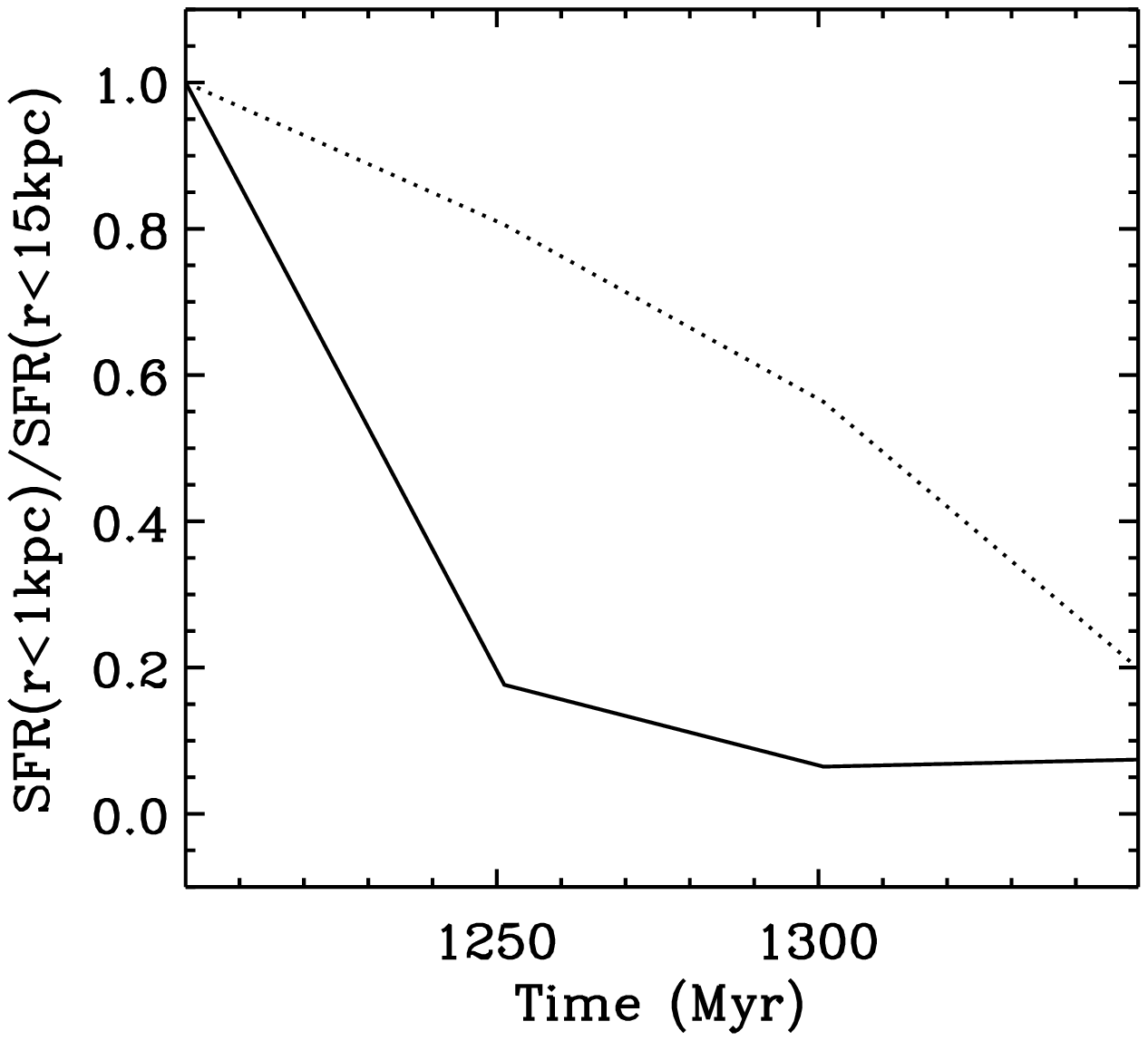}\\
\includegraphics[width=0.32\textwidth]{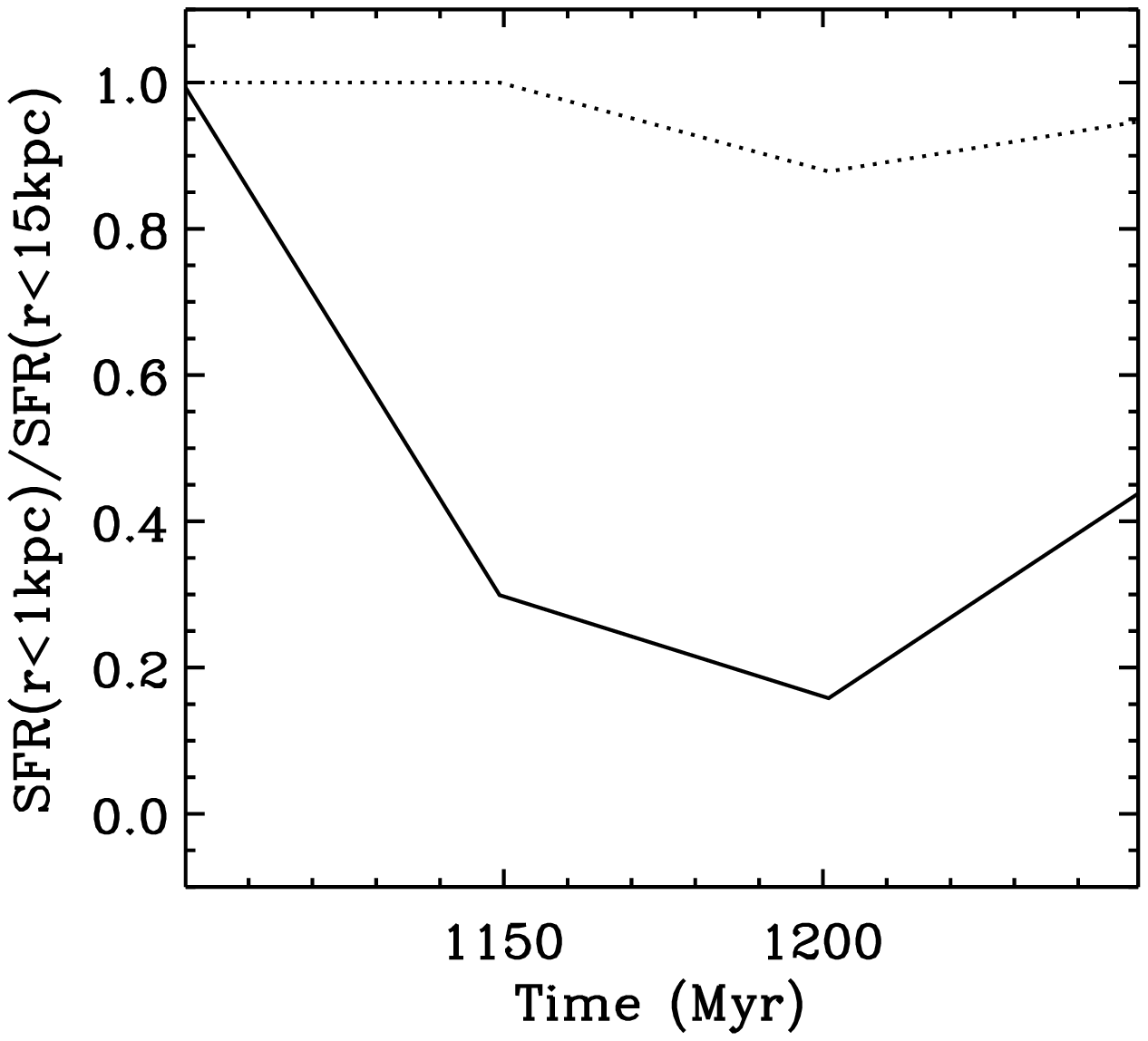}
\includegraphics[width=0.32\textwidth]{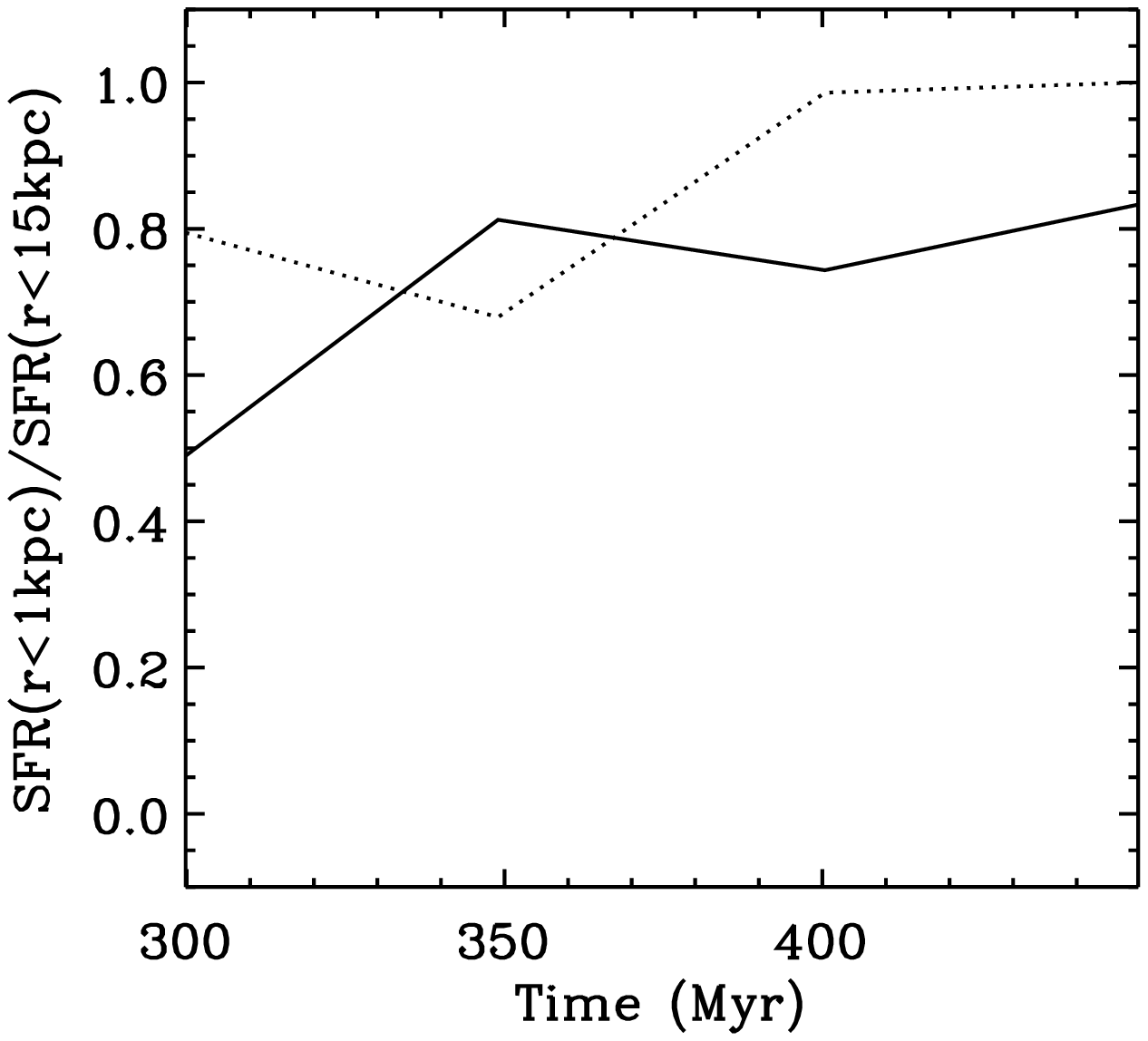}
\caption{Ratio of SFR within $1$kpc of the galaxy centre to SFR within $15$kpc of the galaxy centre for galaxy $1$ (solid line) and galaxy $2$ (dotted line) during the starburst when the two galaxies can still be individually identified. This is defined as when the two galaxy centres are separated by at least $5$ kpc (see text for more details). {\bf From top left to bottom right:} Mergers A, B, C, D and E. }
\label{sfrratio}
\end{figure*}

There is observational evidence for both concentrated nuclear starbursts and the formation of star clusters far from the centre of merging galaxies. We have demonstrated in the previous sections that the mass fraction of dense gas and the velocity dispersion increase significantly as the mergers progress, leading to enhanced star formation. In this section we investigate {\it where} this star formation is taking place.

In Fig.~\ref{sfrratio} we examine the ratio of the SFR within $1$kpc of the galaxy centre to that within $15$kpc of the galaxy centre for galaxy $1$ (solid lines) and galaxy $2$ (dotted lines) in each of the mergers. A ratio of $r\approx1.0$ indicates star formation is centrally concentrated (i.e. something more akin to the classic nuclear starburst is occurring, where the star formation could be smoothly distributed or possibly clumpy), whereas a value of $r<< 1$ indicates that star formation is significantly extended (and, necessarily, clumpy). 

This ratio is only plotted while the two galaxies can be individually identified (by our definition this is when they are at least $5$kpc apart). For the other analyses in this paper, when the galaxies are $< 5$kpc apart, we use the overall centre of mass of the old stars in both galaxies. This is not suitable for studying $r$, because the centre will be in between the nuclei of the two galaxies.  This would introduce the potential to underestimate the value of $r$, since the region of radius $1$kpc around this `combined' centre of mass could contain very little star formation, even if both galaxies have strongly nuclear starbursts.  For this reason, we omit these outputs from Fig.~\ref{sfrratio}. This effect, however, has little impact on the other measurements presented in this paper and, in fact, once reaching a separation of a few kpc, merging galaxies tend to coalesce rapidly (this is also motivation for choosing a threshold of $5$kpc). 

Two of the mergers, C (Fig.~\ref{sfrratio}, top right panel) and D (bottom left panel), have the most extended star formation, with  $r$ reaching below $\approx 0.2$ at times. The mergers B  (Fig.~\ref{sfrratio}, middle right panel) and E  (bottom right panel), have fairly extended star formation, with $r\approx 0.7$. In merger A (Fig.~\ref{sfrratio}, top left panel), the star formation is centrally concentrated (i.e. $r \approx 1$) except around $1200$ Myr when $r\approx 0.4$ for galaxy $1$. 

We can see an illustration of what is happening to cause star formation to become more extended, by returning to the gas density map time sequences in Fig.~\ref{gasdens_timeseq}. In merger D (4th panel) there is a clear change in the distribution of dense gas as the system evolves from $1000$Myr (top left sub-panel), where the gas in both discs is mostly smooth, to $1200$ Myr (top middle sub-panel), where galaxy $1$ is considerably more clumpy. This evolution is reflected in the sfr ratio for galaxy $1$ of D in Fig.~\ref{sfrratio} (bottom left panel, solid line) going from $r=1$ at $t=1000$Myr to $r\sim 0.2$ at $t=1200$Myr. While we do not have a statistical sample, the fact that all mergers have some component of extended star formation that we have demonstrated to be due to the formation/growth of clumps in the gas, which leads to the formation of star clusters, could be an important mechanism to take account of when trying to understand merger-induced star formation.

Fig.~\ref{sfrratio} only covers the time period when the galaxies can be cleanly separated i.e. on the upward curve of the starburst, before its peak. We note that shortly after this time period ends or eventually when the galaxies coalesce, the vast majority of the star formation is concentrated in the inner kpc in all the mergers (i.e. $r$ tends to $1$). However, its not clear that this concentrated starburst is exactly as described in the classic nuclear starburst picture where there is global infall and compression of gas. 

\begin{figure*}
\includegraphics[width=\textwidth]{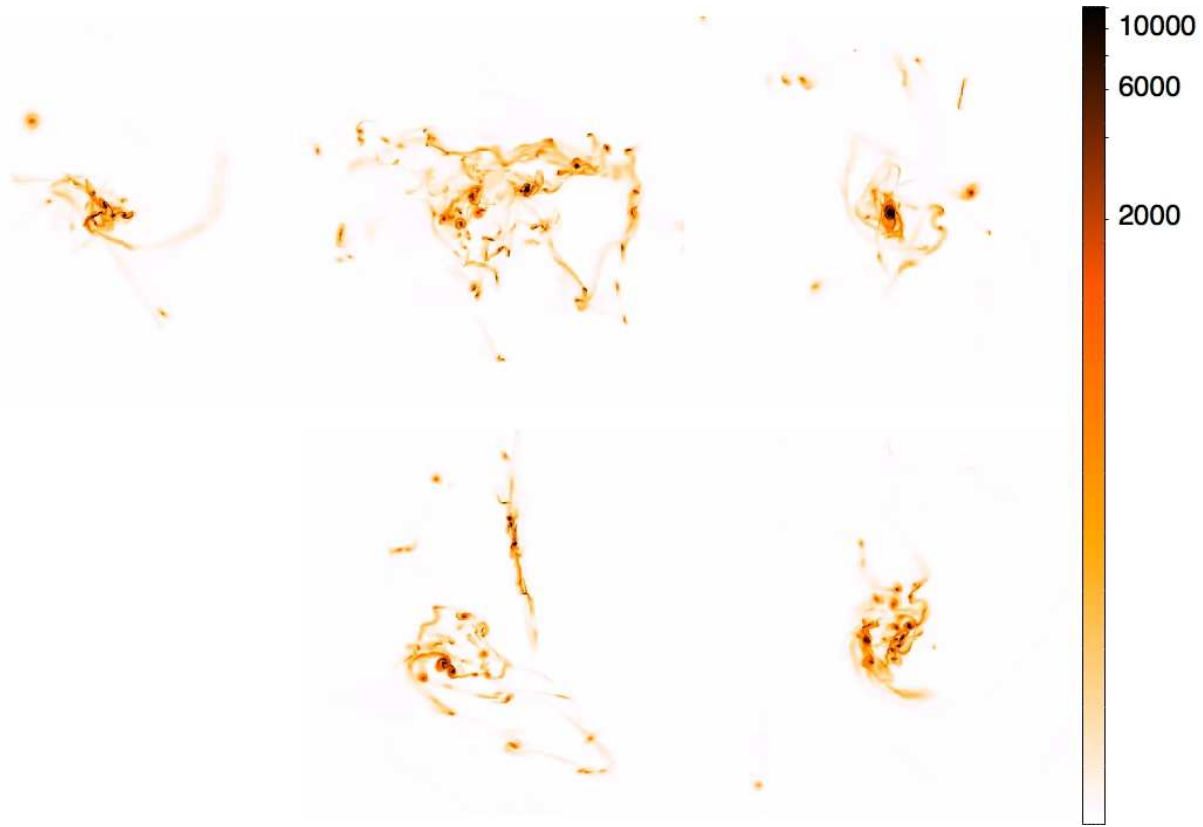}
\caption{Maps showing the maximum gas density along the line of sight in units of H ${\rm cm^{-3}}$ for the central $(4{\rm kpc})^3$ region at the peak of the starburst for each of the mergers.  For ease of comparison between the different mergers, the colour scale is limited at a density of $\sim 10^4$ H ${\rm cm^{-3}}$ in all images. These maps are extracted on the maximum level, level $15$, of the AMR grid, giving them an equivalent resolution of $\approx 5$pc. {\bf From top left to bottom right:} Mergers A, B, C, D and E. }
\label{gasdens_nuclear}
\end{figure*}

In Fig.~\ref{gasdens_nuclear} we show gas density maps for all the mergers at the time of their peak SFR. There is a large variety of gas distributions at the peak SFR in this small sample of mergers. In merger D (bottom left panel), two discs can still be identified by eye indicating that the peak SFR occurs prior to coalescence in this case (although note that according to our identification criteria set out in section \ref{identify}, this is treated as one object in our analysis). Mergers B (top middle panel) and E (bottom right panel) are noticeably clumpy and even though most ($80-90\%$) of the star formation is within $1$ kpc, the stars within this region are still clustered i.e. there is not a large high density core. Mergers A (top left panel) and C (top right panel) have the most centrally concentrated gas distributions (and therefore most centrally concentrated star formation), however  on closer inspection, A is a knot of clumps with tails and only C has a single central star-forming object (and even then there's a few other gas clumps in the vicinity).

We highlight the fact that a centrally concentrated (or `nuclear') starburst can be made up of star clusters; The terms `clustered star formation' and `nuclear starburst' are not mutually exclusive if `nuclear' is assumed to denote only the spatial extent of the star-forming region. Rather there is a difference between {\it nuclear} star formation and {\it extended} star formation. We also note that star formation beyond the nuclear region only seems to occur in the form of clusters i.e. extended star formation {\it is} synonymous with clustered star formation.

In section \ref{origin_vdisp} we demonstrated that a significant increase in velocity dispersion accompanies the increase in SFR during the starburst. It does not, however, seem to be directly correlated with an increase in {\it extended} star formation. For example, in merger D there is a peak in velocity dispersion and SFR (Fig.~\ref{veldisp}, bottom left panel) for galaxy $2$ (blue lines) at $t=1200 {\rm Myr}$. However, if we look at the sfr ratio for D (Fig.~\ref{sfrratio}, bottom left panel) we can see that at the same time galaxy $2$ has much more centrally concentrated star formation $r=1$ than galaxy $1$ ($r=0.2$), the latter of which does not exhibit a similar peak in velocity dispersion.

\section{Interpreting observations: the Kennicutt-Schmidt relation}

\begin{figure*}
\includegraphics[width=0.33\textwidth,trim = 15mm 1mm 7mm 10mm,clip]{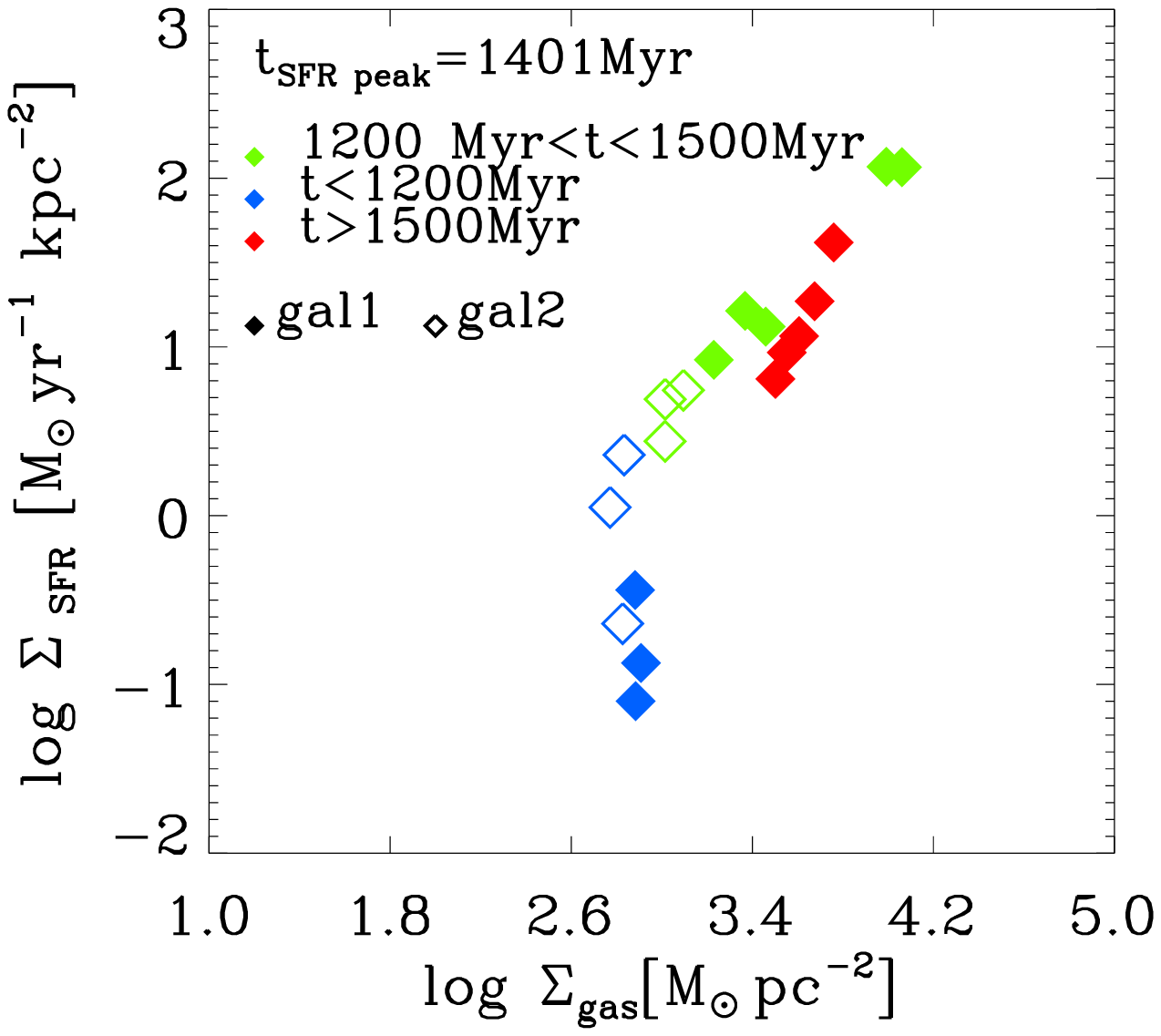}
\includegraphics[width=0.33\textwidth,trim = 15mm 1mm 7mm 10mm,clip]{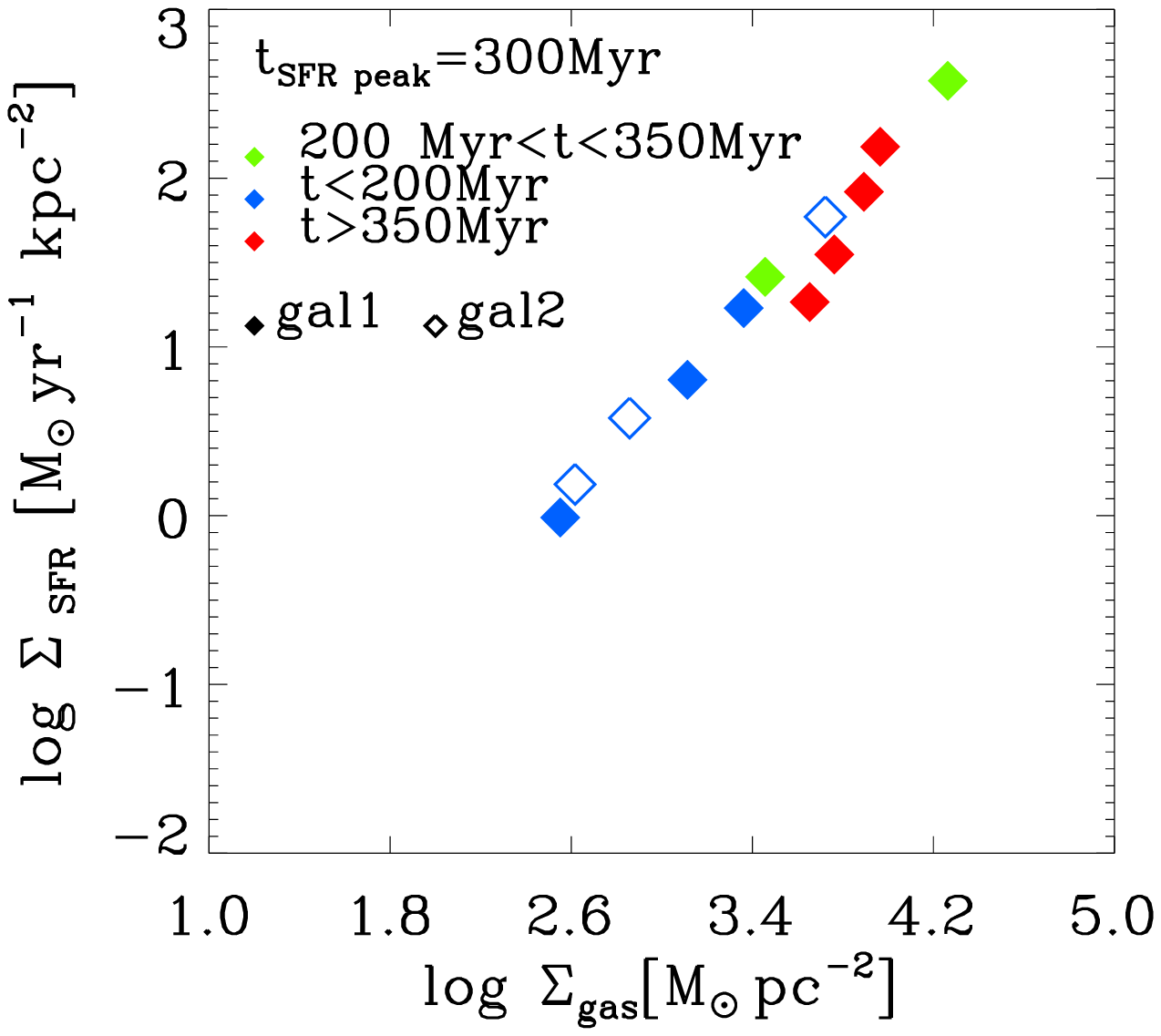}
\includegraphics[width=0.33\textwidth,trim = 15mm 1mm 7mm 10mm,clip]{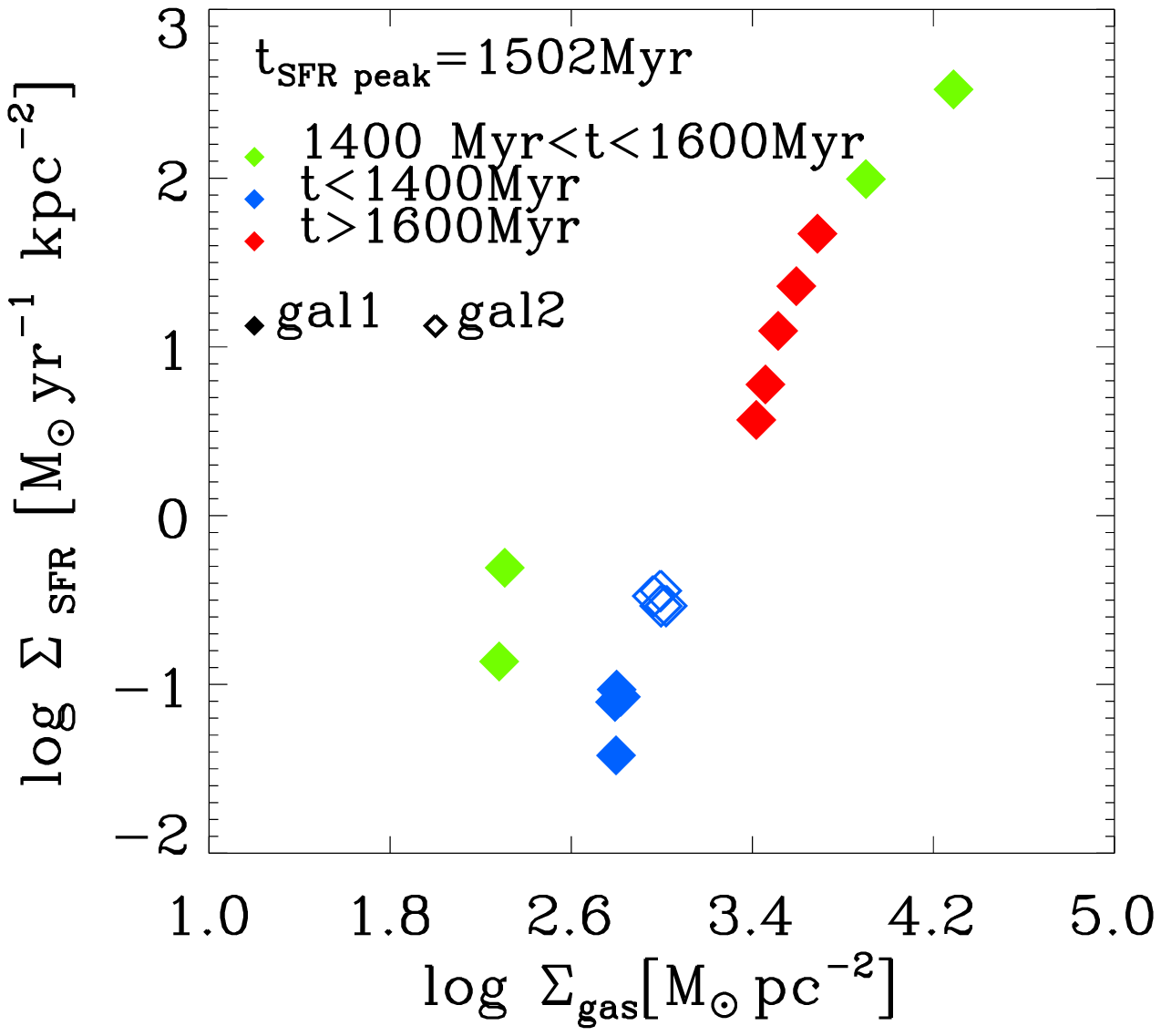}\\
\includegraphics[width=0.33\textwidth,trim = 15mm 1mm 7mm 10mm,clip]{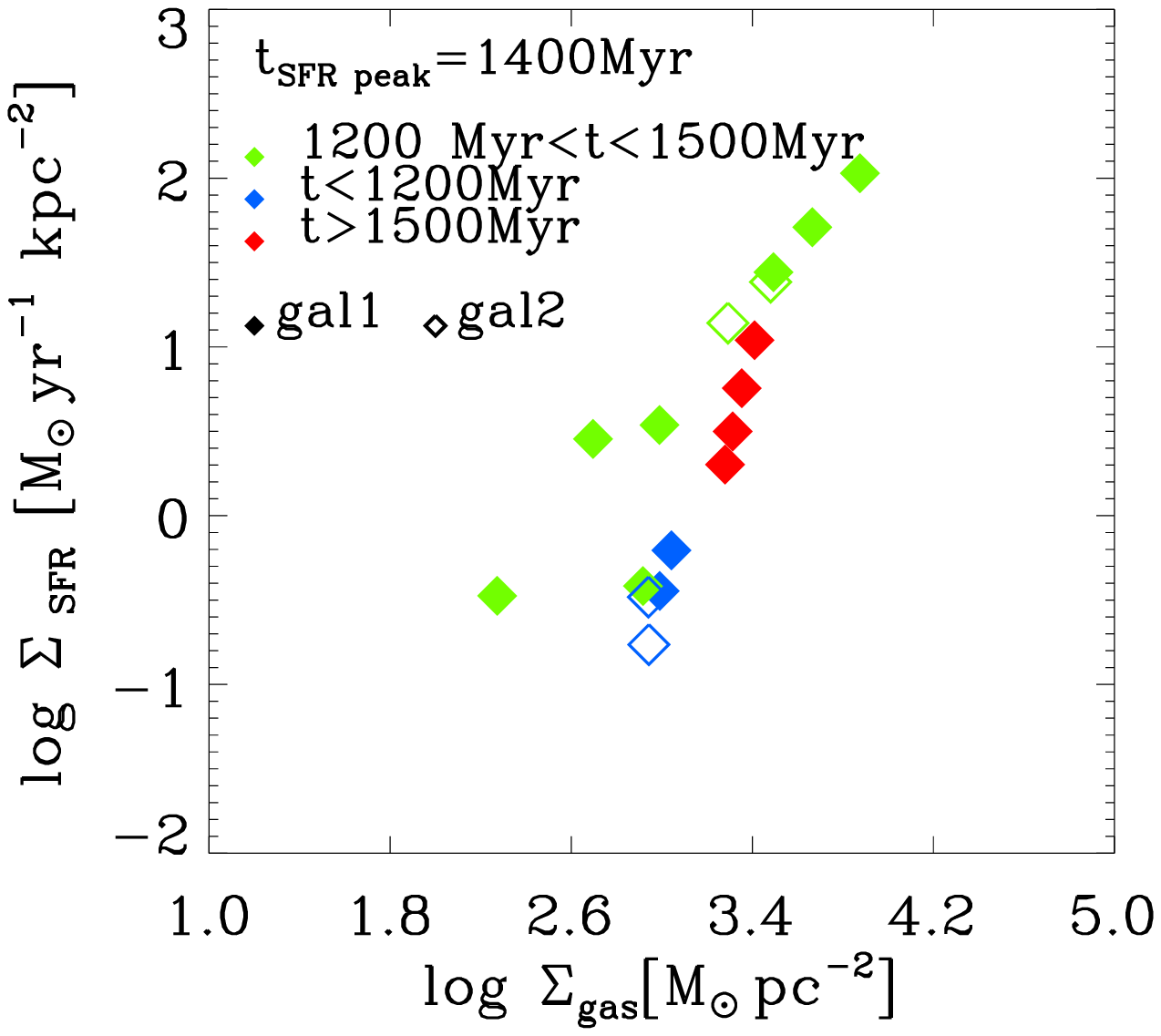}
\includegraphics[width=0.33\textwidth,trim = 15mm 1mm 7mm 10mm,clip]{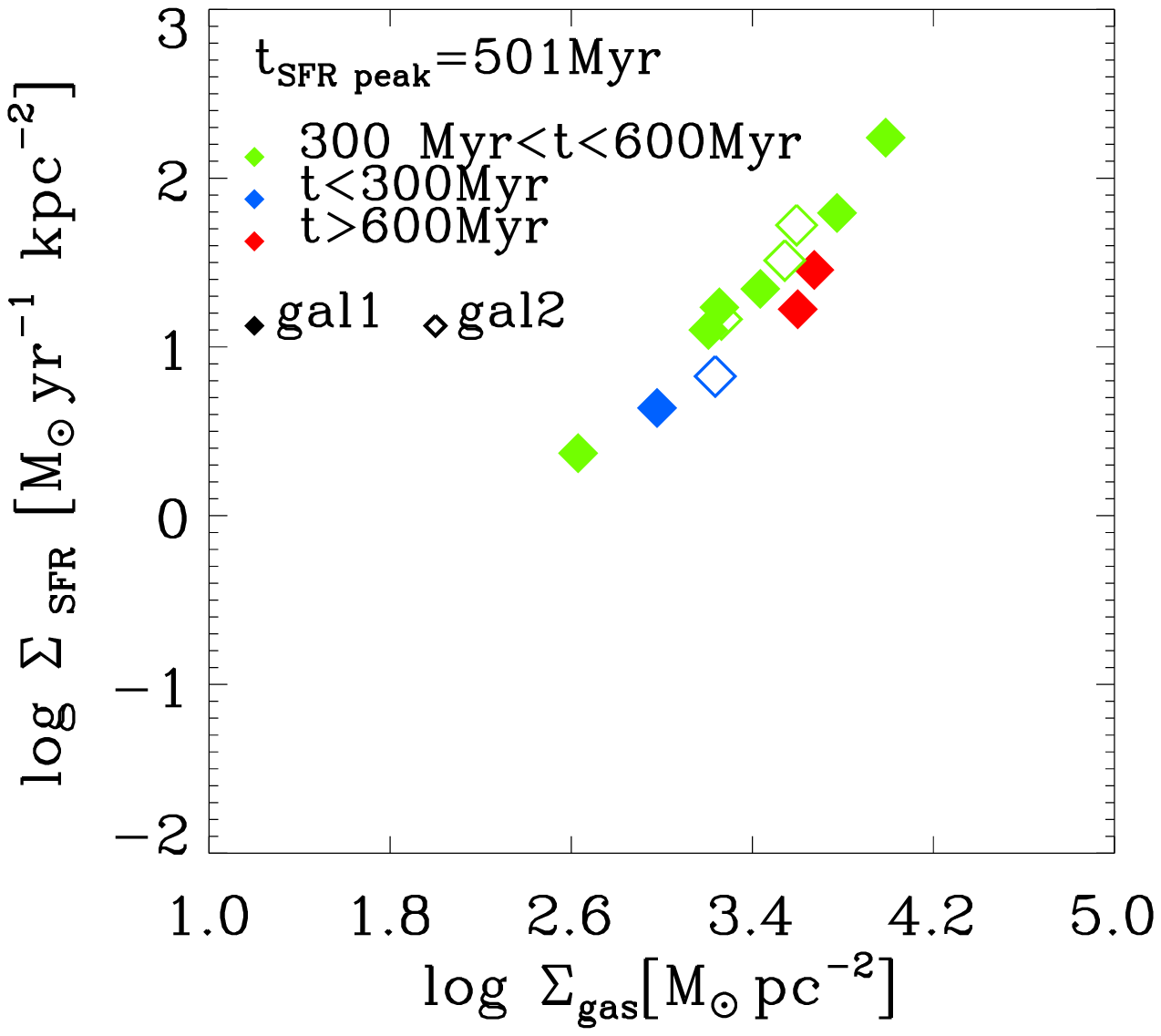}
\includegraphics[width=0.33\textwidth,trim = 15mm 1mm 7mm 10mm,clip]{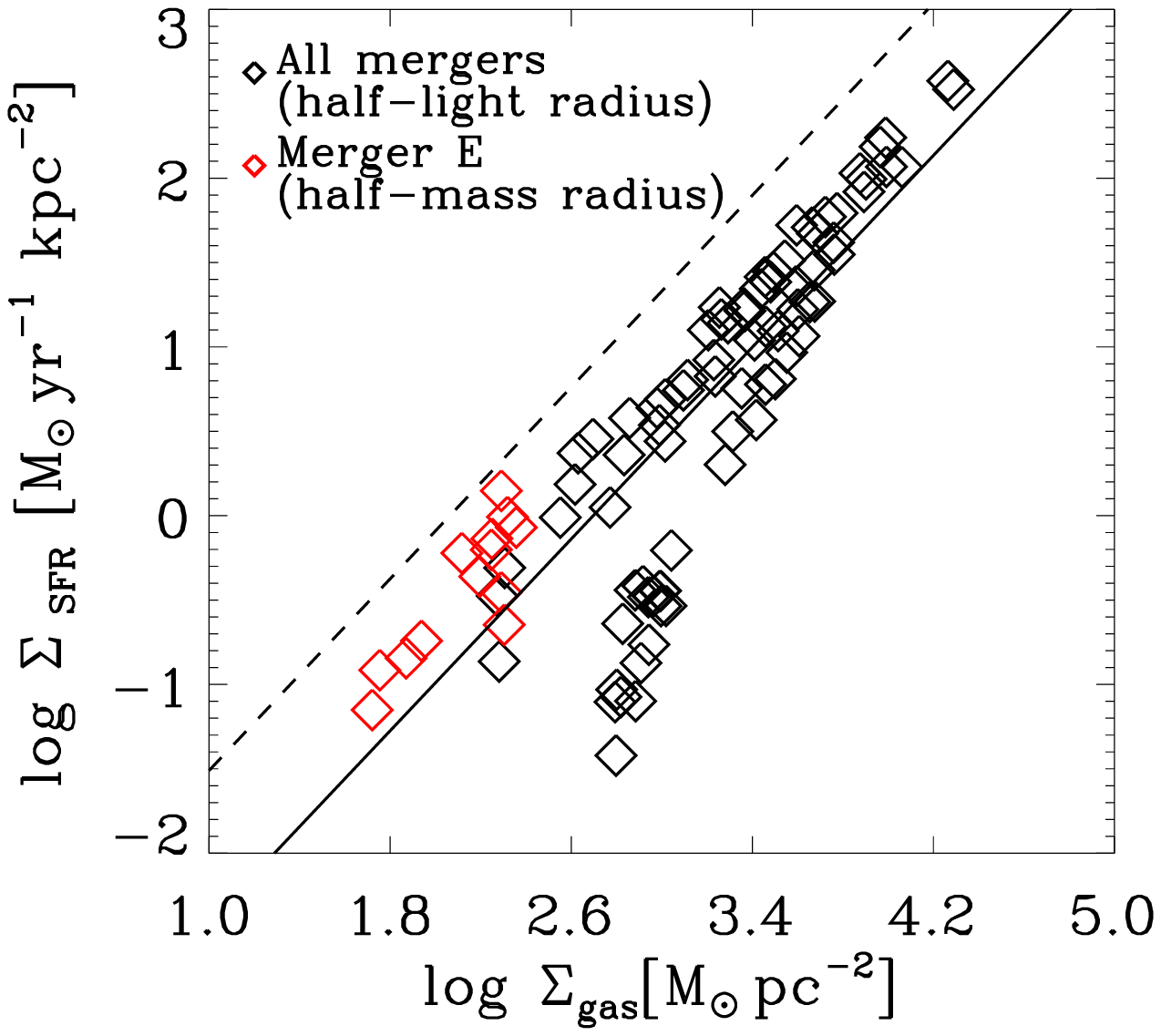}
\caption{Position of galaxy $1$ (filled symbols) and galaxy $2$ (open symbols) on the Kennicutt-Schmidt plot during their interaction, for surface densities computed within the half-light radius. Panels are for merger A (top left), B (top middle), C (top right), D (bottom left), E (bottom middle). Blue points indicate that the galaxies are in the pre-merger phase, on the upward curve of the SFR plot, green points indicate the peak of the starburst and red points indicate the galaxy is in the post-merger phase, on the downward slope of the SFR plot. The points are at $50$ Myr intervals and the phases are chosen by reference to the SFRs in Fig.~\ref{sfr}. The label $t_{\rm SFR peak}$ indicates the time of the output that is closest to the peak of the SFR. The bottom right panel shows the data from all the other panels combined (black symbols) and the points for merger E (red symbols), this time measured within the half-mass radius. Data from \citet{daddi_etal_2010} for the starburst (dashed line) and quiescent (solid line) sequences are over-plotted for reference.}
\label{ks_all}
\end{figure*}

\citet{Kennicutt1998} demonstrated that measurements of SFRs and gas densities in spirals and starburst galaxies were very well fit by a single Schmidt law over several orders of magnitude in both quantities. In recent years the global nature of the Kennicutt-Schmidt star formation relation has been re-examined. \citet{daddi_etal_2010}  and \citet{GenzelTacconiGracia-Carpio2010} show that the data can also be fit by two Schmidt laws with different normalisations: one for starbursts and one for quiescent discs. This offset disappears, however,  if the gas surface densities are divided by the dynamical time, suggesting the SFR is correlated to global galaxy properties rather than being universal.  \citet{antennae_ramses} show that in simulations of the Antennae system that can reproduce the observed star cluster formation, the system does indeed move from the quiescent to the starburst sequence. This suggests that extreme merger-induced clustered star formation (not resolved in many previous merger simulations) could provide a physical origin for the two star formation sequences.

 \citet{SaintongeTacconiFabello2012} compare a sample of galaxies, selected to be evenly distributed in the $M_{\star}$-SFR parameter space, to a subset that has the same distribution in the parameter space as a volume-limited sample. The former inherently has an excess of high specific SFR galaxies and very massive galaxies. They find that when examining the unbiased subset, no bimodality in the Kennicutt-Schmidt relation is found because the extremely efficient star-forming mergers that give rise to the starburst sequence in the full sample \citep[and in previous work, e.g.][]{daddi_etal_2010, GenzelTacconiGracia-Carpio2010} are in fact rare objects and have little influence overall. They find that more `average' mergers are typically offset above the mean Kennicutt-Schmidt relation.

There are also significant uncertainties in the conversion factor, $\alpha_{\rm CO}$, used to extrapolate from the observed CO emission, to the mass of ${\rm H}_2$. This conversion factor has been shown to be different for starbursts and spirals, so different values are used for the various galaxy populations \citep[e.g.][]{SolomonVanden-Bout2005}. \citet{NarayananKrumholzOstriker2011} use hydrodynamical simulations  of discs and mergers  and radiative transfer in order to compute an accurate fitting formula for $\alpha_{\rm CO}$, based on local conditions in a wide range of systems. The conversion factor varies smoothly (with metallicity and CO line intensity) and using this to construct a Kennicutt-Schmidt plot (rather than several discreet values for the different galaxy populations) results in a universal star formation law. 

Except at very low metallicity, for a given mass of ${\rm H}_2$, there is a relatively constant mass of CO, but this CO emits more light if its transitions are excited by collisions (among other things), which are more frequent in denser environments. The gas density PDFs for our simulated mergers (see Fig.~\ref{pdf_all}) show clearly that there is an increasing excess of dense gas produced during the merger. Therefore, our simulations also predict that there should be more CO emission coming out of the system as the merger progresses and thus a lower $\alpha_{\rm CO}$, compared to that in quiescent discs.

Alternatively, \citet{KrumholzDekelMcKee2011} claim that the observations of different star formation laws for different objects is a projection effect, caused by frequent discrepancies between column density and local, 3D density. They show that a local, {\it volumetric} star formation law holds for a wide range of observations. In summary, it is still unclear whether there is a bimodality or simply a range of values in the Kennicutt-Schmidt relation and particularly whether merger-induced star cluster formation could potentially explain this.

Given that we resolve star cluster formation in our merger simulations and we have chosen fairly `average' orbital parameters for a $\Lambda$CDM universe, we can add a new perspective on the Kennicutt-Schmidt relation to those from existing theoretical work. We calculate the half-light radius by assigning a luminosity to the star particles according to their mass and age \citep{WeidnerKroupaLarsen2004} ,

\begin{eqnarray}
L (a< 10 {\rm Myr})&\propto &M_{\rm star} \nonumber \\
L (a>10{\rm Myr})  & \propto & M_{\rm star} \left( \frac{a}{10{\rm Myr}} \right) ^{-0.7} \nonumber
\end{eqnarray}

\noindent For stars in the initial conditions, which have ages equal to zero by definition when the simulation starts, we draw their ages at the start at random from the range $0-5$Gyr. For a given output, $\Sigma_{\rm SFR}$ is calculated within the half-light radius using the SFR averaged over the previous $10$ Myr. For simplicity, we compute the half-light radii in 3-dimensions and use all gas and stars within this volume for the calculations of $\Sigma_{\rm gas}$ and $\Sigma_{\rm SFR}$, however the quantities are divided by the area $\pi r_{3D}^2$. 

Fig.~\ref{ks_all} shows the time evolution of  galaxy $1$ (filled diamonds) and galaxy $2$ (open diamonds) in each of the mergers on the Kennicutt-Schmidt plot (from top left to bottom middle: mergers A, B, C, D and E). To highlight the behaviour as the mergers progress, the points (which are at $50$ Myr intervals) have been colour-coded as pre-merger discs (blue symbols), galaxies near/at the peak of the starburst (green symbols) and post-merger galaxies (red symbols). The peak starburst phases are determined by eye with reference to the SFRs in Fig.~\ref{sfr} and are intended merely as a guide. We see a general trend whereby the galaxies move towards the right as the merger progresses (i.e. the green points are mostly to the right of the blue points), as expected if the gas is undergoing global compression. The galaxies also move upwards as they enter the peak of the starburst (i.e. the green points are typically higher than the blue points). Towards the end of the starburst (the red points), the galaxies then move back down and to the left as they become `red and dead'. 

It is the vertical motion of the galaxies in the Kennicutt-Schmidt plot that is most interesting as this could be related to the starbursting sequence (vertically above the quiescent disc sequence) seen in observations \citep{daddi_etal_2010, GenzelTacconiGracia-Carpio2010}. In Fig.~\ref{ks_all} (bottom right panel) we plot the data for all the mergers on the Kennicutt-Schmidt plot (black symbols), with the best-fitting quiescent sequence (solid line) and starburst sequence (dotted line) from \citet{daddi_etal_2010} over-plotted for reference. We note that the absolute position of our simulated galaxies in the Kennicutt-Schmidt plot is due to calibration of the parameters controlling star formation to give appropriate SFRs (our isolated discs lie on the quiescent sequence -see section \ref{sf_compare} for further details). The relative movement in the position of the points as the merger progresses, however, is indicative of changes in the physical behaviour of the gas. There is not a clear bimodality for our simulated mergers as most of the points lie slightly above the quiescent sequence. This is still compatible with the results of \citet{daddi_etal_2010} and \citet{GenzelTacconiGracia-Carpio2010} as these studies select extreme starburst galaxies with much higher SFRs ($\sim 100 {\rm M}_{\odot} {\rm yr}^{1}$) than our simulated mergers ($\sim 10 {\rm M}_{\odot} {\rm yr}^{-1}$). It is possible, therefore, that it is simply a selection effect causing the apparent bimodality i.e. the starburst sequence is an upper limit, reached by the most star-forming systems, and more average mergers lie somewhere in between this and the quiescent sequence. This is supported by the observations of \citet{SaintongeTacconiFabello2012} (discussed in more detail at the beginning of this section). Our finding that merging galaxies typically lie slightly above the quiescent sequence and that the (bulge-dominated) remnants lie below it is in good agreement with \citet{SaintongeTacconiFabello2012}.

\citet{antennae_ramses} found the same general pattern in Kennicutt-Schmidt plots of the Antennae system, with two important differences (we refer here to their $12$pc high resolution run). Firstly, they measured a much more significant vertical jump (from the quiescent sequence towards the starburst sequence) during the merger and, secondly, the galaxies remained above the quiescent sequence after the merger had ended. The difference in evolution after the merger can be explained by the lack of SN feedback in the Antennae simulations, which is remedied in the simulations presented here. In the former case, once a dense gas clump has formed it is not possible to destroy it allowing star formation to continue unhindered, whereas feedback can limit this process.

Understanding the difference in the evolution from the quiescent to the starburst sequence is more complex as there are several possible explanations. It is possible that the SN feedback helps to regulate the amount of clustered star formation, by dispersing gas clumps when the SFR reaches a certain level i.e. that the contribution from clumpy star formation is naturally limited in our simulations and the galaxies cannot get as close to the starburst sequence without this. In this scenario, it is possible that the importance of clumpy star formation is overestimated in \citet{antennae_ramses} i.e. that work demonstrates the upper limit of its potential impact. 

Alternatively, the Antennae could be considered an unusual system in the sense that the orbital parameters maximise the amount of clustered star formation (after all, it is one of the key examples of super star cluster formation). For more `average' orbital parameters (as chosen intentionally in this study), the clustered star formation is possibly just less significant. We also point out that due to the orbital parameters of the Antennae system, the galaxies undergo a relatively long period of enhanced clustered star formation away from the nucleus, during the time between first and second pericenter ($\approx 200$)Myr. In the mergers presented here, the galaxies coalesce much sooner after the start of the starburst and so have less time to form star clusters, before the gas is driven inward.

An interesting feature of the Kennicutt-Schmidt plots in Fig.~\ref{ks_all} is that the galaxies in mergers A (top left panel), C (top right panel) and D (bottom left panel) start significantly below the quiescent sequence, whereas the galaxies in B (top middle panel) and E (bottom right panel) start {\it on} the quiescent sequence. This difference arises because mergers B and E occur rapidly, but the galaxies in A, C and D take much longer to merge simply because of their different orbital parameters (note the different timing of the SFR peaks in Fig.~\ref{sfr}). This means that the galaxies in the former cases have  evolved off the quiescent sequence (they have much lower SFRs and are typically much less clumpy) before merging. We note that there {\it is} a significant vertical jump towards the starburst sequence in these cases. In these examples, as in the Antennae simulation, the pre-merger discs are not clumpy (one can either consider this as a resolution issue, or simply modelling a smoother disc). It is possible then that a smooth disc will move towards the starburst sequence when the interaction boosts the clump Jeans mass above the resolution limit, but if the clumps in the disc are already resolved (like merger B and E) this effect is not seen. It is not yet clear if this could be a physical process, or a result of the inevitable limits on the minimum clump mass that can be resolved in simulations.

It is also possible that the way of measuring the Kennicutt-Schmidt values (both in simulations and observations) is not robust. \citet{McQuinnSkillmanDalcanton2012} show that starburst dwarf galaxies exhibit both extended and concentrated star formation but that using short timescale star formation tracers can lead to those with extended star formation not being classified as starbursts. Essentially, the choice of technique can bias the measurements to favour what is happening in the nuclear region. With this in mind, we consider whether the way in which we make the Kennicutt-Schmidt measurements may affect the results, since the half light radii are small (typically $\sim 0.5$kpc) in our simulations. 

We recompute the Kennicutt-Schmidt measurements for our mergers within the $80$ per cent light radius and within the half gas mass radius. For the former there is no discernible change, but for the latter (which is $\sim 2-3$kpc) some of the points for some of the mergers are slightly closer to (though still not on) the starburst sequence and, as expected, all the points move down and to the left since the densities are lower when average over a larger area.  In Fig.~\ref{ks_all} (bottom right panel) we show the Kennicutt-Schmidt measurements for merger E within the half-mass radius (red points), to demonstrate this behaviour. The effect is considerably stronger in merger E than in the other mergers, suggesting that the dependence on the radius used is related to the specific distribution of the gas and stars. The conclusions we have drawn are not affected by changing the radius within which we measure the Kennicutt-Schmidt parameters, but it is worth noting that the absolute position on the Kennicutt-Schmidt plot is somewhat sensitive to this and that some measures may be biased towards detecting nuclear star formation.

\section{Discussion: Is clustered extended star formation in mergers important?}

Our study has shown that the appearance of a clustered, extended component of merger-induced star formation is common and not restricted to special systems like the Antennae. In the earlier stages of the merger, there is a phase of extended, clustered star formation (where the length of the phase is dependent on the orbital parameters). This boosts the SFR to a moderate degree, but occurs prior to the peak of the starburst.  The merging galaxies then go through a very short phase, in which the peak SFR is reached, which is dominated by the nuclear starburst (i.e. star formation occurring within $1$kpc of the centre) driven by global gas compression. We note, however, that during the nuclear starburst phase there is still evidence of fragmentation. It is not clear overall which mode of star formation dominates the star formation budget. Although the nuclear starburst produces the peak SFR, this is for a very short time period, whereas extended, clustered star formation has a much longer phase. 

We have demonstrated that while our merger simulations do not support the idea of a marked bimodality in the the Kennicutt-Schmidt relation, this mode of clustered star formation can explain offsets from the quiescent disc sequence. Studying the clustered component is particularly important if we wish to understand the distribution and ages of stars formed during mergers. The stars formed in this mode are formed earlier, over a longer time period and up to several kpc from the centre of the system. Correctly simulating/resolving the clustered component of star formation in mergers is also crucial if we are to try to explain the globular cluster populations, for which merger-induced cluster formation is one of the proposed formation channels \citep[e.g.][]{KruijssenInti-PelupessyLamers2011}.

\section{Conclusions}

In this paper we simulate $5$ equal-mass galaxy mergers with AMR code {\sc ramses} in order to investigate the mechanisms for merger-induced star formation. With $\approx 5$pc spatial resolution in the densest regions, we capture the multiphase nature of the ISM and can resolve clumpy star formation. The aim is to establish if the extended, clustered star formation observed in systems like the Antennae galaxies may also be an important feature of more `average'  mergers. Our main findings are as follows.

\begin{itemize}

\item We observe significant evolution in the density PDFs as the mergers progress, resulting in an excess of very dense gas. This provides an explanation for the enhanced HCN/CO ratios observed in ULIRGs. We find that our proxies for the luminosity ratios increase with increasing SFR as seen in the observations of \citet{JuneauNarayananMoustakas2009} and find good agreement in the best fit slopes for this correlation.

\item This growing excess of dense gas means our simulations also predict that there should be increasing CO emission as the merger progresses (due to enhanced probability of collisional excitation etc in a denser environment) while the ${\rm H}_2$ mass remains constant, resulting in a lower $\alpha_{\rm CO}$ compared to that in quiescent discs.

\item We find that the starburst is also accompanied by a peak in the average 1D velocity dispersion, $\sigma_{1\rm D}$. The value of $\sigma_{1\rm D}$ increasing from $\approx20$km/s in the isolated galaxy to up to $\approx 80$km/s at the peak of the starbursts in the mergers. We have confirmed the increased velocity dispersion is not a result of the SN feedback and therefore must result from the interaction itself.

\item The mergers exhibit a variety of distributions of star-forming regions, from concentrated in the central kpc to spread over several kpc. All the mergers have a component of extended (i.e. beyond the central kpc), clustered star formation at some point during the early stages of the starburst, but star formation becomes nuclear as the galaxies approach coalescence. Extended star formation is always clumpy and we also note that, in some cases, even when star formation is within $1$ kpc, the stars in this region are still clustered. The formation of star clusters is important, therefore, whether the star formation is extended or concentrated.

\item We do not see a clear bimodality in the Kennicutt-Schmidt plot for pre-merger and merging galaxies, but rather a range of values between the two sequences of \citet{daddi_etal_2010}. 

\end{itemize}

\section*{Acknowledgments}

We thank the referee, Desika Narayanan for a positive, constructive report. We also thank Eric Emsellem, Pierre-Alain Duc and Emanuele Daddi for numerous helpful discussions and Sadegh Khochfar for useful suggestions. We acknowledge support from the French Agence Nationale de la Recherche under contract ANR-08-BLAN-0274-02 (P.I. Eric Emsellem), and from the EU through grant ERC-StG-257720 and the CosmoComp ITN.
The simulations were performed at the TGCC computing center as part of a GENCI project (grants 2011-042192 and 2012-042192).

\bibliographystyle{mn2e_mw} 
\bibliography{merger}
 \label{lastpage}

\end{document}